\documentclass[11pt]{article}          
\usepackage[top=1in, bottom=1in, left=1in, right=1in]{geometry}
\usepackage{graphicx}
\usepackage{amsfonts,amssymb,amsmath,amsopn}
\usepackage{graphics}
\usepackage{rotating}
\usepackage{bigints}
\usepackage{booktabs}
\usepackage{float}
\usepackage{subfigure}
\usepackage{todonotes}
\usepackage{url}
\usepackage{amssymb}
\usepackage{color}
\usepackage{bm}
\usepackage{multirow}
\usepackage{diagbox}
\usepackage{enumerate}
\usepackage{epstopdf}
\usepackage[ruled]{algorithm2e}
\usepackage[colorlinks=true]{hyperref}
\usepackage{lineno}
\usepackage{amssymb}
\usepackage{booktabs}
\usepackage[misc]{ifsym}

\newtheorem{asu}{{\sc Assumption}}
\newtheorem{thm}{Theorem}
\newtheorem{prop}{Proposition}
\newtheorem{lem}{Lemma}
\newcommand{\bfsigma}{\bm{\sigma}}
\newcommand{\bfx}{\mathbf{x}}

\newcommand{\bfv}{\mathbf{v}}
\newcommand{\bfy}{\mathbf{y}}
\newcommand{\bfW}{\mathbf{W}}
\newcommand{\bfR}{\mathbf{R}}
\newcommand{\Bistar}{B_{i_\star}}
\newcommand{\bfxtilde}{\tilde{\mathbf{x}}}
\newcommand{\bfxtildep}{\tilde{\mathbf{x}}^{\text{p}}}
\newcommand{\bfxtildeo}{\tilde{\mathbf{x}}^{\text{o}}}
\newcommand{\bfxtildel}{\tilde{\mathbf{x}}^{\text{l}}}
\newcommand{\bfxp}{\bfx^{\text{p}}}
\newcommand{\bfxo}{\bfx^{\text{o}}}
\newcommand{\bfxl}{\bfx^{\text{l}}}
\newcommand{\Phil}{\Phi^{\text{l}}}
\newcommand{\red}{}
\newcommand{\black}{}
\newcommand{\blue}{\textcolor{blue}}

\newcommand{\mb}{\mathbf}
\newcommand{\E}{\mathbf{E}}

\newtheorem{remark}[thm]{Remark}

\begin{document}

\title{Accelerating Metropolis-within-Gibbs sampler with localized computations of differential equations
}


\author{Qiang LIU \footnote{Department of Mathematics, National University of Singapore, Singapore, matliuq@nus.edu.sg}        \and
       Xin T. TONG \footnote{Department of Mathematics, National University of 
           Singapore, Singapore, mattxin@nus.edu.sg}
}


\maketitle

\begin{abstract}
Inverse problem is ubiquitous in science and engineering, and Bayesian methodologies are often used to infer the underlying parameters. 
For high dimensional temporal-spatial models, classical Markov chain Monte Carlo (MCMC) methods are often slow to converge, and it is necessary to apply Metropolis-within-Gibbs (MwG) sampling on parameter blocks.  
However, the computation cost of each MwG iteration is typically $O(n^2)$, where $n$ is the model dimension. This can be too expensive in practice. 
This paper introduces a new reduced computation method to bring down the computation cost to $O(n)$, for the inverse initial value problem of a stochastic differential equation (SDE) with local interactions. 
The key observation is that each MwG proposal is only different from the original iterate at one parameter block, and this difference will only propagate within a local domain in the SDE computations. 
Therefore we can approximate the global SDE computation with a surrogate updated only within the local domain for reduced computation cost. 
Both theoretically and numerically, we show that the approximation errors can be controlled by the local domain size. 
We discuss how to implement  the local computation scheme using Euler--Maruyama and 4th order Runge--Kutta methods. 
We numerically demonstrate the performance of the proposed method with the Lorenz 96 model and a linear stochastic flow model.

\end{abstract}

\section{Introduction}
\subsection{Inverse problem and MCMC}
Inverse problem is ubiquitous in various fields of science and engineering \cite{W1996,CSN1977,KS2005}.
It concerns how to infer model parameters from partial, delayed, and noisy observations. 
Typical examples include using measurements of seismic waves to determine the location of an earthquake epicenter, and recovering images that are as close to natural ones as possible from blurry observations \cite{R2016}. The general formulation of inverse problem can be written as 
\begin{equation}
\label{geninv}
\bfy =h(\bfx,\zeta).
\end{equation}
Here $\bfx$ denotes the parameters to be estimated, $\bfy$ denotes the observation data, $h$ is a physical model describing the data collection process, and $\zeta$ is the possible random factor involved. 

Often, it is of interest to quantify the uncertainty of $\bfx$, which can be used to infer estimation accuracy and regulate risk. The Bayesian approach is more appropriate for this purpose \cite{KS2005,CDRS2009,S2010}. This involves modeling $\bfx$ with a prior distribution $p_0(\bfx)$ and finding the observation distribution $p_l(\bfy|\bfx)$ from \eqref{geninv}. Then the \black{Bayes'} formula indicates that the posterior distribution of $\bfx$ is given by 
\begin{equation}\label{post_dis}
p(\bfx|\bfy)  \propto p_0(\bfx) \cdot p_l(\bfy|\bfx).
\end{equation}

Markov chain Monte Carlo (MCMC) is a big class of stochastic algorithms designed to sample the posterior density iteratively \cite{GRS1996,L2003}. In each iteration, a new proposal $\bfx'$ is generated from the current iterate $\bfx$, which can be described by a transition density $Q(\bfx'|\bfx)$. Then the Metropolis-\black{Hastings} (MH) step accepts this proposal with probability 
\begin{equation}\label{genMH}
\alpha(\bfx',\bfx)=\min\left\{1,\frac{Q(\bfx|\bfx')p_0(\bfx') p_l(\bfy|\bfx')}{Q(\bfx'|\bfx)p_0(\bfx)  p_l(\bfy|\bfx)}\right\}.
\end{equation}
Some popular choices of $Q$ include random walk transition in random walk Metropolis (RWM), and  Langevin dynamic transition in Metropolis adjusted Langevin algorithm (MALA) \cite{MRRTT1953,H1970,RR1998}.

While these MCMC algorithms perform well for classical problems, they become very slow when applied to modern data science problems, where the parameter dimension $n:=\text{dim}(\bfx)$ is very large. The main issue is that the proposal $\bfx'$ in RWM or MALA  in general is different from $\bfx$ at all components, and the MH-acceptance probability $\alpha(\bfx',\bfx)$ is often of order $O(e^{-\|\bfx'-\bfx\|^2})\approx O(e^{-n})$. This is extremely small when  $n$ is more than a few thousands. The proposals will mostly be rejected, and the MCMC is essentially stuck. One way to alleviate this issue is to choose a small step size in the \black{proposal}, so the average acceptance probability is $O(1)$. But then the consecutive MCMC iterates are close to each other, so the overall movement of MCMC can still be slow. 

\subsection{Spatial localization and MwG }
The curse of dimensionality can often be lifted if there are exploitable statistical \black{structures}. Examples include conditional Gaussianity and low effective dimension. In geophysical applications, the components of $\bfx$ usually describe status at different locations. 
Because the underlying physical law is often \black{short-ranged}, faraway components of $\bfx$ are nearly independent. 
Consequentially, the associated covariance matrix will have a clear banded \black{structure}. 
Such \black{a phenomenon} is called \emph{spatial localization}, and it widely exists in problems involving vast spatial domains.
In \black{the statistic literature}, such banded \black{structure} can be exploited by tapering techniques, which significantly \black{improve} covariance estimation.
And in numerical weather prediction (NWP), localization techniques are designed to utilize this \black{structure}, so algorithms such as ensemble Kalman filter can provide stable estimation for planetary models of $10^8$ dimensions with merely 100 samples.

A recent work \cite{MTM2019} investigates the possibility to exploit spatial localization with MCMC. It is found that Gibbs sampling  \cite{GG1984,GS1990} is a natural framework for this purpose. To do so, one first partitions the model components into $m$ blocks with $\bfx=[\bfx_1,\ldots, \bfx_m]$, where each block contains only $b=n/m=O(1)$ nearby components. When running \black{MCMC}, one generates a proposal for the $k$-th block by applying Gibbs sampler on the prior. This proposal $\bfx'$ will be different from $\bfx$ only at the $k$-th block, which is of dimension $b$. Therefore, the corresponding MH acceptance probability $\alpha(\bfx',\bfx)$ in \eqref{genMH} will be \black{of scale $O(1)$} and does not degenerate even if the overall dimension $n$ is large. A completely new iterate can be generated by repeating this procedure for all $m$ blocks. A more detailed description of this Metropolis-within-Gibbs (MwG) sampler can be found in Section \ref{sec:MwG}. Numerical tests and rigorous analysis in Gaussian settings have revealed that MwG has dimension independent MCMC convergence rate when the underlying distributions are spatially localized. 

\subsection{Acceleration with local computation}
While MwG takes only a constant number of iterations to sample the posterior distribution, the computation cost of each iterate can be expensive. This is when a Gibbs block proposal is being processed by the MH step \eqref{genMH}, one often needs to evaluate $p_l(\bfy |\bfx')$. \black{It often involves $O(n)$ computational cost}. The proposal procedure is repeated for all $m$ blocks. So, to generate a new MwG iterate, the computation cost is $O(n)\times m=O(n^2)$. \black{This is much more expensive than the ones for RWM and MALA iterates, which in general cost $O(n)$ computation}.

However, it is possible to reduce the cost of  $p_l(\bfy|\bfx')$ to $O(1)$. The main observation is that $\bfx'$ and $\bfx$ differ only at one block, say the $k$-th block, and the computation of $p_l(\bfy|\bfx)$ is done in the previous MH step. So if one can replace the $\bfx_k$ part in the computation of $p_l(\bfy|\bfx)$ with $\bfx'_k$, the value of $p_l(\bfy|\bfx')$ can be obtained cheaply. As a simple example, suppose in \eqref{geninv} the observation model is $\bfy=\bfx+\xi$ with $\xi\sim\mathcal{N}(\mathbf{0}, \mathbf{I}_n)$, a normal distribution with $n$-dimensional mean vector $\mathbf{0}$ and covariance matrix $\mathbf{I}_n$ being an $n \times n$ identity matrix. Then, we have 
\[
-\log p_l(\bfy|\bfx)\propto \sum_{k=1}^m \|\bfy_k-\bfx_k\|^2. 
\]
Suppose this value is already available, then when computing $-\log p_l(\bfy|\bfx')$, one only needs to update the $k$-th block in the summation to $\|\bfy_k-\bfx'_k\|^2$, which only costs $O(1)$ \black{computation}. This example can be easily generalized to  cases where $\bfy_k$ relies on multiple blocks of $\bfx$. See detailed discussion of this and the possibility of parallelization in \cite{MTM2019}.

In this paper, we explore the possibility of reducing the computational cost of MwG from $O(n^2)$ to $O(n)$. We assume the observation model \eqref{geninv} is given by a high dimensional stochastic differential equation (SDE) with short-range interaction, where $\bfx=\bfx(0)$ is the initial condition of the SDE, and $\bfy$ consists of noisy partial observations of the SDE, $\bfx(s\leq T)$, in a fixed time interval $[0,T]$. Such an inverse initial value problem is practically important. It can be interpreted as an one-step smoothing problem in signal processing. Data assimilation problems such as NWP can also be formulated as sequential applications of it \cite{CR2011,MH2012}. 

Finding $p_l(\bfy|\bfx)$ is equivalent to solving the SDE and finding $\bfx(s\leq T)$ in the smoothing context. Standard Euler--Maruyama scheme would require a \black{computational} cost of $O(n)$. When the Gibbs sampler proposes $\bfx'$, one needs to use it as a new SDE initial condition and compute $\bfx'(s\leq T)$. But because $\bfx'$ is different from $\bfx$ only for $\bfx_k$,  we show in Proposition \ref{bound_opz} that $\bfx_i'(s\leq T)$ is not much different from $\bfx_i(s\leq T)$ if $|i-k|$ is larger than a radius $L$. Therefore it is not necessary to do the re-computation in a full way. In other words, instead of applying Euler--\black{Maruyama} to compute $\bfx_i'(s\leq T)$ for all $i\in\{1,\ldots, m\}$, we only compute locally for the ones with $|i-k|\leq L$. Since the radius $L$ can often be chosen as a constant independent of $n$, updating $\bfx(s\leq T)$  to $\bfx'(s\leq T)$ only \black{costs $O(1)$ computation}. This procedure will be repeated through all $m$ blocks, so the overall cost of finding a new MwG iterate is $O(1)\times m=O(n)$. This achieves the aforementioned computation reduction objective. We use a-MwG to denote this accelerated version of MwG.

Since MwG only requires constantly many iterates to sample a spatially localized distribution, a-MwG is expected to solve the Bayesian inverse problem with only a computation cost of $O(n)$. This is the optimal dimension scaling one can obtain for any numerical methods. The main drawback of a-MwG is that it uses a local computation scheme, so it is subjective to approximation errors. However, these errors can be controlled by choosing a large enough $L$. We demonstrate this through rigorous analysis and numerical tests. 

\subsection{Organization and \black{Preliminaries} }
\label{pre}
This paper is organized in the following way. In Section \ref{sec:setup}, we consider the inverse problem of how to infer the initial \black{condition} of an SDE with local \black{interaction}. We review the MwG sampler with a discussion of its computational complexity. We derive an accelerated-algorithm, called a-MwG sampler, in Section \ref{sec:acc}, and analyze the approximation errors. We also discuss how to  implement  a-MwG with Euler--Maruyama and 4th order Runge--Kutta schemes, as well as its adaptation to parallelization. In section \ref{sec:exp}, two numerical experiments of the Lorenz 96 and linearized stochastic model are studied. The paper is concluded in Section \ref{sec:con}. All the proofs are postponed to the Appendix.

Throughout this paper, we will use the following notations. When applying MwG, we need to partition a high \black{dimensional} vector into blocks, for which we write as $\bfx=[\bfx_1,\ldots, \bfx_m]$. For simplicity of the discussion, we assume each block shares the same length $b$, so the total dimension $n=mb$. We remark our result is easily generalizable to non-constant block sizes. In practice, each block often \black{represents} information at a location on a torus, therefore it is natural to introduce measure of distance between indices as $d(j_1,j_2) = \min \{ |j_1-j_2|, |j_1-j_2+m|, |j_1-j_2 + m|\}$ with $j_1, j_2 = 1,...,m$.

When a matrix $A$ is given, the $(i,j)$-th entry is denoted as $A_{i,j}$.  $A^{T}, A^{-1}$ denote the transpose and the inverse of the matrix respectively. We adopt $\|\cdot \|$ and $\|\cdot\|_{\infty}$ to denote the $l_2$ norm and $l_\infty$ norm for a vector, namely, for a vector $\vec{a}$ with elements $a_1,...,a_n$, we have $ \| \vec{a} \| = \sqrt{\sum_{i=1}^{n}a_i^2}$, and $\|\vec{a}\|_{\infty} = \max\{ |a_i|: i=1,...,n \}$. For an $m \times n$ matrix $A$, the $l_2$ operator is written as $\|A\| = \sup{\{ \|A\vec{v}\|: \vec{v} \in \mathbb{R}^{n}, \|\vec{v}\|=1 \}}$. For two $m\times n$ matrices $A$ and $B$ (including vectors as a special case), we say $A\preceq (\succeq) B$ if we have $A_{i,j} \leq (\geq) B_{i,j}$ for $1 \leq i \leq m$, $1 \leq j \leq n$ entry-wise. $\mathbf{0}_{m \times n}$ and $\mathbf{1}_{m \times n}$ are $m\times n$ matrices whose entries are 0 and 1 respectively. $\mathcal{N}(\mathbf{m}, \Sigma)$ is a multidimensional normal distribution with mean vector $\mathbf{m}$ and covariance matrix $\Sigma$. We use $\mathbf{I}_n$ to denote $n \times n$ identity matrix.  $C$ is a generic positive constant that may varies from line to line.

\section{Problem setup}\label{sec:setup}
\subsection{Inverse initial value problem for SDE}
We consider a spatial-temporal model with local interaction 
\begin{align}\label{sys:original}
\begin{split}
& d\bfx_j(t) = \mathbf{f}_j(t, \bfx_{j-1}(t), \bfx_j(t),\bfx_{j+1}(t) )dt \\
&\qquad \qquad + \bm{\sigma}_j(t,\bfx_j(t))d\bfW_j(t), \ \text{for} \ j=1,...,m, \\
& \bfx_{0}(t) = \bfx_{m}(t), \ \bfx_{m+1}(t) = \bfx_{1}(t), \  t \in [0,T],
\end{split}
\end{align}
where $\bm{\sigma}_j$ is an $b \times b$ \black{matrix-valued} adapted locally bounded process, and $\bfW_j(t)$ is an \black{$b$-dimensional} standard Brownian motion. We assume the following Lipschitz continuity conditions are satisfied for the coefficient processes $\mathbf{f}_j$ and $\bm{\sigma}_j$:
\begin{asu}\label{lip}
	Given $b \times 1$ vectors $\bfx_{1}, \bfx_{2}, \bfx_{3}$, $\bfy_{1}, \bfy_{2}, \bfy_{3}$ and $t \in [0,T]$, for $j=1,...,m$, there exists constants $C_{\mathbf{f}}>0$ and $C_{\bm{\sigma}}>0$ such that
	\begin{align*}
	& \| \mathbf{f}_j(t, \bfx_{1}, \bfx_{2}, \bfx_{3} ) - \mathbf{f}_j(t, \bfy_{1}, \bfy_{2}, \bfy_{3} ) \|^2 \\
	&  \leq C_{\mathbf{f}}(\|\bfx_{1}-\bfy_{1} \|^2 + \|\bfx_{2}- \bfy_{2}\|^2 + \|\bfx_{3}-\bfy_{3}\|^2 ),
	\end{align*}
	and 
	\begin{align*}
	&\|(\bm{\sigma}_j(t,\bfx_1) - \bm{\sigma}_j(t,\bfy_1))  \|^2 \leq C_{{\bfsigma}}\|\bfx_1-\bfy_1\|^2.
	\end{align*}
\end{asu}

Assumption \ref{lip} is widely used to guarantee the solution to \eqref{sys:original} exists and is unique, when the initial condition and the realizations of the Brownian motion $\bfW = [\bfW_1, ..., \bfW_m]$ are given. We write the solution as $\bfx(t)=\Phi(\bfx(0),t,\bfW)$. In the scenario where there is no stochastic forcing, namely $\bfsigma_j\equiv \mathbf{0}_{b \times b}$, \eqref{sys:original} is an ordinary differential equation (ODE). Its solution can be simply written as $\bfx(t)=\Phi(\bfx(0),t)$.

One key \black{feature} of \eqref{sys:original} is that the drift term driving $\bfx_j$ relies only on its neighboring blocks. This describes general short-range interactions that are typical in spatial-temporal models. The formulation of \eqref{sys:original} can be naturally derived, if one considers applying finite difference discretization for a stochastic partial differential equation, such as the reaction-diffusion equations. Details of such derivation can be found in \cite{CMT18cam}.

We assume an $n'$-dimensional data is generated from noisy observation of \eqref{sys:original} at  time $T>0$, that is
\begin{align}\label{inv}
\bfy = \mathbf{H}\cdot\Phi(\bfx,T,\bfW) + \mathbf{\xi}. 
\end{align}
Here $\mathbf{H}$ is an $n'$ by $n$ observation matrix that serves as selecting the observable components, and $\xi$ represents the associated observation noise, which is  distributed as $ \mathcal{N}(\mathbf{0}_{n' \times 1},\bfR)$.

The Bayesian inverse problem this paper trying to solve is finding the posterior distribution of the initial condition $\bfx$ with a given data $\bfy$. 
For simplicity, we assume the prior distribution $p_0$ of $\bfx$ is Gaussian with mean $\mathbf{0}_{n \times 1}$ and covariance matrix $\Sigma_{\text{pri}}$. Then the observation likelihood is given by 
\begin{equation}
\label{eqn:obslike}
p_{l}(\bfy|\bfx)\propto \E\exp(-\tfrac12\|\bfy-\mathbf{H}\cdot\Phi(\bfx,T,\bfW)\|^2_{\bfR}),
\end{equation}
where for a vector $\mathbf{v}$, we define $\|\mathbf{v}\|^2_\bfR:=\bfv^T\bfR^{-1}\bfv$. 
The expectation in $\eqref{eqn:obslike}$ is averaging over all \black{realizations} of $\bfW$. 
In computation, it can be approximated by a Monte Carlo sampled version 
\begin{equation}
\label{eqn:obsmc}
\hat{p}_{l}(\bfy|\bfx):=\frac1{\black{S}}\sum_{c=1}^{\black{S}}\exp(-\tfrac12\|\bfy-\mathbf{H}\cdot\Phi(\bfx,T,\bfW^{(c)})\|^2_{\bfR}),
\end{equation}
where each $\bfW^{(c)}$ is an independent Brownian motion realization. \black{Note that \eqref{eqn:obsmc} is an unbiased estimator of \eqref{eqn:obslike}, using it instead of \eqref{eqn:obslike} in  MCMC is known as the pseudo-marginal algorithm \cite{AG09}. The convergence of the pseudo-marginal algorithm is largely similar to the standard MCMC where \eqref{eqn:obslike} is used directly \cite{AV15}. }

When \eqref{sys:original} is an ODE, \eqref{eqn:obslike} is simplified as 
\begin{equation}
\label{eqn:ODElike}
p_{l}(\bfy|\bfx)\propto \exp(-\tfrac12\|\bfy-\mathbf{H}\cdot\Phi(\bfx,T)\|^2_{\bfR}).
\end{equation}

\subsection{MwG and its computational complexity}
\label{sec:MwG}
The Metropolis-within-Gibbs sampler is an MCMC algorithm. For each block, it generates a proposal by applying Gibbs sampling to the prior, and use the Metropolis step to incorporate data information. In specific, it generates iterations of form  $\bfx^{k} = [\bfx^{k}_1, ..., \bfx^{k}_m]$ through the following steps, where we start with $k=1$ and draw $\bfx^{1}$ randomly from $p_0$:
\begin{enumerate}[1)]
	\item Repeat steps 2-4  for all block index $j=1,\ldots,m$
	\item Sample $\bfxtilde_j$ from
	\[
	p_0(\bfx_j\in \,\cdot\,|\bfx^k_{1},\ldots, \bfx^k_{j-1}, \bfx^k_{j+1},\ldots, \bfx^k_m).
	\]
	\item Let $\bfxp=[\bfx^k_{1},\ldots, \bfx^k_{j-1}, \bfxtilde_j, \bfx^k_{j+1},\ldots, \bfx^i_m]$.
	\item Let $\bfx^k=\bfxp$ with probability 
	\begin{equation}
	\label{eqn:MH}
	\alpha(\bfxp,\bfx^k)=\min\left\{1, \frac{p_l(\bfy|\bfxp)}{p_l(\bfy|\bfx^k)}\right\}.
	\end{equation}
	$p_l$ can also be replaced by the sample version $\hat{p}_l$ in  \eqref{eqn:obsmc}.
	\item When the loop in step 1) is finished, let $\bfx^{k+1}=\bfx^k$, and increase \black{the iteration}  index from $k$ to $k+1$.  
\end{enumerate}
In \cite{MTM2019}, it is shown that MwG has dimension independent performance if 1) $p_0$ is a Gaussian distribution, and its covariance or precision matrix is close to be banded; 2) each component of $\bfy$ has significant dependence only on a few components of $\bfx$. It is also discussed how to truncate the far-off diagonal entries of the prior covariance or precision matrix. This \black{so-called} ``localization" technique can simplify the computation of the proposal probability in MwG step 2), making the cost to be $O(1)$. 

When applying MwG directly to the inverse problem described in Section \ref{sec:setup}, the computational cost for fully updating all blocks is $O(n^2)$. This is because in step 4), we need to evaluate $p_l$ through either \eqref{eqn:obsmc} or \eqref{eqn:ODElike}. This involves finding the numerical solution to \eqref{sys:original}, which requires executing Euler--Maruyama or 4th order Runge--Kutta methods. Both these methods require $O(n)$ computation complexity when the dimension of the differential equation \eqref{sys:original} is $n$. Then because step 4) is repeated for all $m$ blocks, the total computation cost is $O(n)\times m=O(n^2)$. 

In summary, although the vanilla MwG only requires O(1) iterates to converge to the posterior distribution, each individual iterate can cost $O(n^2)$ \black{computation}. This can be less appealing than standard MCMC algorithms with optimal tuning. For example, RWM can converge to the posterior distribution with $O(\sqrt{n})$ iterations, while each iterate costs $O(n)$ complexity, so the total complexity is $O(n^\frac32)$. In this paper, we demonstrate how to bring down the computational cost of each MwG iterate to $O(n)$. This will lead to the optimal  MCMC computational scalability.

\section{Acceleration with local computation}\label{sec:acc}
\subsection{Spatial propagation of local changes}\label{def_xop}
Based on our previous discussion, the main computational cost of MwG takes place at step 4) when \eqref{eqn:MH} is evaluated.
We will discuss how to reduce this cost to $O(1)$. In what follows, We fix the \black{MCMC iteration} index as $\text{o}$ and block index as $i_\star$, while the same procedure applies to all \black{iteration} indices and blocks. 

First of all, it should be noted that when executing step 4),  the value of $p_l(\bfy|\bfxo)$ is already available from the previous \black{iteration} step 4). Likewise, we already have the values of $\bfxo(t\leq T)=\Phi(\bfxo, t\leq T, \bfW)$. It is only necessary to find $p_l(\bfy|\bfxp)$, 
or equivalently $\bfxp(t\leq T)=\Phi(\bfxp, t\leq T, \bfW)$. Note that we will often write $\bfxo(0)$ as $\bfxo$, since it is the information we try to recover. 

Our main observation here is that $\bfxp(0)=\bfxp$ is different from $\bfxo(0)=\bfxo$ only at the $i_\star$-th block. 
Since components in SDE \eqref{sys:original} exchange information only through local interactions, 
the differences between $\bfxp(t)$ and $\bfxo(t)$ are likely to be of significance only for blocks that are close to $i_\star$. 
In fact, we have the following proposition:

\begin{prop}\label{bound_opz}
	Under Assumption \ref{lip} and with  any given positive constant $C_d$, for $j=1,...,m$, we have  
	\begin{align}\label{thm1_res1}
	\begin{split}
	&\E[\|\mb{x}^{\text{\emph{o}}}_j(t) - \mb{x}_j^{\text{\emph{p}}}(t)\|^2] \\
	&\leq 	\|\mb{x}^{\text{\emph{o}}}_{i_\star}-\mb{x}^{\text{\emph{p}}}_{i_\star}\|^2e^{C_1(\mathbf{f},\bm{\sigma})t }  e^{-C_d \cdot d(j,i_{\star})}
	\end{split}
	\end{align}
	with 
	\begin{align}\label{C_1}
	C_1(\mathbf{f},\bm{\sigma}) = (e^{C_d}+e^{-C_d}+1)(C_{\mathbf{f}}+C_{\bfsigma}+1).
	\end{align}
\end{prop}

In particular, for any fixed small enough threshold $\epsilon$ and time range $T$, we can find a radius $L$, such that 
\[
\E[\|\bfxo_j(t) - \bfxp_j(t)\|^2] \leq \epsilon, \quad \forall \ t\leq T,\ \black{d(i_\star, j) > L}. 
\]
We define the \emph{local domain centered at index $i_\star$} as 
\[
\Bistar:=\{j : d(i_\star, j)\leq L\},
\]
and use  $\Bistar^c$ to denote its complement. Then $\bfxo_j(t)$ is already a good approximation of $\bfxp_j(t)$ for $j\in \Bistar^c$. 
In other words, it is no longer necessary to recompute $\bfxp_j(t)$ for $j\in \Bistar^c$, and we only need to compute $\bfxp_j(t)$ for $j\in \Bistar$.
Since $\Bistar$ contains at most $2L+1$ elements, and $L$ is likely to be independent with the problem dimension $n$\black{. This} provides us with a way to accelerate the overall computation of MwG. 

\subsection{Local surrogate}\label{def_xlp}
\label{sec:local}
Next we consider how to use existing values of $\bfxo(t\leq T)$ to reduce the computation of $\bfxp(t\leq T)$. 
For this purpose, we introduce a local surrogate model given by
\begin{align}\label{sys:local}
\begin{split}
&  \bfxl_j(t)  =  \bfx^{\text{o}}_j(t), \ \text{for} \ j \in \Bistar^c,\\
& d\bfxl_j(t) = \mathbf{f}_{j}(t,\bfxl_{j-1}(t), \bfxl_j(t),\bfxl_{j+1}(t) )dt  \\
& \qquad \qquad + \bm{\sigma}_{j}(t,\bfx_j^{\text{l}}(t))d\bfW_j(t), \ \text{for} \ j \in \Bistar,\\
& \bfxl_{0}(t) = \bfxl_{m}(t), \ \bfxl_{m+1}(t) = \bfxl_{1}(t). 
\end{split}
\end{align}
Its initial condition is set to be $\bfxl(0)=\bfxp(0)=\bfxp$. We write its solution as 
\[
\bfxl(t)=\Phil(\bfxp, t,\bfxo(t\leq T), \bfW).
\] 
Note that the local surrogate $\bfxl_j(t)$ within $\Bistar$ depends on $\bfxo_j(t)$ for the boundary blocks of $\Bistar^c$, of which the indices are usually just $i_\star+L+1$ and $i_\star-L-1$. 
Such dependence can be viewed as using $\bfxo$ as spatial boundary conditions for the computation of $\bfxl$. 

We will use the local surrogate $\bfxl(t\leq T)$ as an approximation of $\bfxp(t\leq T)$.
It can be computed cheaply because it is different from $\bfxo(t\leq T)$ only for blocks inside the local domain $\Bistar$. 
The details of how to achieve this can be found in subsequent parts. First of all, we investigate the approximation error through the following theorem 
\begin{thm}\label{thm_con}
	Under Assumption \ref{lip} and with  any given positive constant $C_d$, for $j=1,...,m$ and $t\in [0,T]$, we have  
	\begin{align}\label{thm1_res2}
	\begin{split}
	&\E[ \|\bfx^{\text{\emph{l}}}_j(t) - \mb{x}_j^{\text{\emph{p}}}(t)\|^2 ]\\
	& \leq C_2(\mathbf{f}, \bfsigma) \| \mb{x}^{\emph{o}}_{i_\star}-\mb{x}^{\emph{p}}_{i_\star} \|^2 e^{2C_1(\mathbf{f}, \bfsigma) t} e^{-C_d\cdot(L+1)},
	\end{split}
	\end{align} 
	with $C_1(\mathbf{f}, \bfsigma)$ defined in (\ref{C_1}) and 
	\begin{align}\label{C_2}
	C_2(\mathbf{f}, \bfsigma)= \max{ \Big\{ \dfrac{ 2C_{\mathbf{f}} }{C_1(\mathbf{f}, \bfsigma)}, 1 \Big\} }. 
	\end{align}
	In particular, for any given $\epsilon>0$, if the local domain radius $L$ satisfies
	\begin{align}\label{thm1_res3}
	L \geq \dfrac{\log{\Big( \dfrac{\epsilon}{C_2(\mathbf{f}, \bfsigma) \|\mb{x}^{\text{\emph{o}}}_{i_\star}-\mb{x}^{\text{\emph{p}}}_{i_\star}\|^2} \Big)}}{-C_d}   +\dfrac{2C_1(\mathbf{f},\bm{\sigma})}{C_d}T, 
	\end{align}
	then $\E[ \|\bfx^{\text{\emph{l}}}_j(t) - \mb{x}_j^{\text{\emph{p}}}(t)\|^2 ]\leq \epsilon$ for all $t\leq T$. 
\end{thm}
Theorem \ref{thm_con} indicates that we can use $\bfxl(T)$ to approximate $\bfxp(T)$ with controlled accuracy. So, $\alpha(\bfxp,\bfxo)$ as defined in (\ref{eqn:MH}) can be substituted by $\alpha(\bfxl,\bfxo)$. When such a scheme for calculating (\ref{eqn:MH}) is plugged into MwG, this will lead to acceleration in terms of computation complexity. We call it accelerated \black{Metropolis-within-Gibbs} (a-MwG) and present its pseudocode in Algorithm \ref{alg:acc_l_mwg}. Next, we will discuss how to compute the local surrogate \eqref{sys:local} with numerical methods, and how to implement parallelization.  

\begin{algorithm}[!htbp]\label{alg:acc_l_mwg}
	\caption{The accelerated Metropolis-within-Gibbs sampling}
	\KwIn{$K$: number of iterations; $b$: block size; $L$: local domain radius; $\black{S}$: Brownian motion sample size}
	\KwOut{$K$ iterations $\bfx^{k}$ with $k=1,\ldots, K$. }
	\blue{\%Initial computation} \\
	Sample $\bfx^{0}\sim p_0$;\\	
	\For{ $c = 1,...,\black{S}$}{
		Generate Brownian motion $\bfW^{(c)}(t\leq T)$;\\	
		Let $\bfxo(t\leq T,c) = \Phi(\bfx^0, t\leq T, \bfW^{(c)})$;\\
		\blue{\%Full model computation}
	}
	Let
	$\hat{p}^{\text{o}}_{l}=\frac1{\black{S}}\sum_{c=1}^{\black{S}}\exp(-\tfrac12\|\bfy-\mathbf{H}\cdot \bfxo(T,c)\|^2_{\bfR})$;\\
	\blue{\%MCMC loop}\\
	\For{ $k = 1,...,K$}{
		\blue{\%Gibbs loop}	\\
		\For{$j = 1,...,m$}{
			Sample\qquad\qquad \blue{\%Proposal step}\\
			$\bfxp_{j} \sim p_0( \bfx_j\in \,\cdot\,| \bfxo_1, ..., \bfxo_{j-1}, \bfxo_{j+1}, ..., \bfxo_m) $\;
			Let $\bfxp = [\bfxo_1, ..., \bfxo_{j-1},\bfxp_j, \bfxo_{j+1}, ..., \bfxo_m]$\;
			\For{ $c = 1,...,\black{S}$}{
				Let $\bfxl(t,c) = \Phil(\bfxp, t,\bfxo(t\leq T,c), \bfW^{(c)})$ \\ for all $t\leq T$.  \blue{\%Local computation} }			 
			\blue{\%\black{pseudo-marginal }MH step}\\
			Let $\hat{p}^{\text{p}}_{l}=\frac1{\black{S}}\sum_{c=1}^{\black{S}}\exp(-\tfrac12\|\bfy-\mathbf{H}\cdot \bfxl(T,c)\|^2_{\bfR})$;\\
			Let $\alpha = \min\{ 1, \hat{p}^{\text{p}}_{l}/\hat{p}^{\text{o}}_{l}\}$;\\ 
			Sample $u \sim \text{Uniform}([0,1])$\;
			\If{$u \leq \alpha $}{
				Let $\bfxo=\bfxp$;\\
				Let $\hat{p}^{\text{o}}_{l}=\hat{p}^{\text{p}}_{l}$;\\
				\For{ $c = 1,...,\black{S}$}{
					Let $\bfxo(t\leq T,c) = \bfxl(t\leq T ,c )$;
			}} 
		}
		Let $\bfx^k = \bfxo$\;
	}
\end{algorithm}

\subsection{Local computation with Euler--Maruyama}\label{Euler}
When stochastic forcing is nonzero, our model \eqref{sys:original} is a bona-fide SDE. 
Euler--Maruyama is the standard numerical method for its computation. 

In the vanilla MwG, if one applies it directly to the full model \eqref{sys:original} to obtain 
$\bfxp(T)$, a small step size $h>0$ will be chosen, and the value of $\bfxp(t)$ will be approximated \black{by $\bfxtildep((i+1)h)$ if $ih < t \leq (i+1)h$.} The numerical solution \black{$\bfxtildep((i+1)h)$ at grid points $(i+1)h$} is generated by the following iterations, starting from $\bfxtildep(0)=\bfxp$:
\begin{align}\label{tilxp}
\begin{split}
\bfxtildep_j((i+1)h)&=  \bfxtildep_j(ih)  +\bfsigma_j(ih,\bfxtildep_j(ih))\sqrt{h}W_{i,j} + \\
& \quad \mathbf{f}_j(ih,\bfxtildep_{j-1}(ih), \bfxtildep_j(ih),\bfxtildep_{j+1}(ih) )h. 
\end{split}
\end{align}
Here, $W_{i,j}$ are independent samples from $\mathcal{N}(\mathbf{0}_{b\times 1}, \mb{I}_b)$. One can view 
\[
\bfW_j(kh)=\sum_{i=1}^{k} \sqrt{h}W_{i,j}
\]
as a realization of the Brownian \black{motion} $\bfW_j$ in \eqref{sys:original}. As (\ref{tilxp}) is repeated for all block index $j$, obtaining $\bfxtildep(T)$ has an $O(n)$ computational complexity.

When applying Euler--Maruyama  for the local surrogate \eqref{sys:local}, we assume numerical approximation of $\bfxo(t\leq T)$ are available as $\bfxtildeo(ih)$. 
This comes either from the previous a-MwG iteration or an initial computation before the first a-MwG iteration. Then if the proposal $\bfxp$ is different from $\bfxo$ at the $i_\star$\black{-th} block, we set the local domain as $\Bistar=\{j: d(j,i_\star)\leq L\}$, and denote its complement as $\Bistar^c$. For blocks outside $\Bistar$, we directly use $\bfxtildeo_{j}(ih)$ as the numerical solution, that is we let
\begin{equation}
\label{eqn:EMbc}
\bfxtildel_j(ih)=\bfxtildeo_{j}(ih),\quad j\in \Bistar^c, \ i=1,\ldots, T/h.
\end{equation}
New computation is needed for $\bfxtildel_j(ih)$ with $j\in \Bistar$, which are obtained through the following 
\begin{align}\label{eqn:EMloc}
\notag
&\bfxtildel_j((i+1)h)=  \bfxtildel_j(ih)  +\bfsigma_j(ih,\bfxtildel_j(ih))\sqrt{h}W_{i,j} \\
& \quad \quad + \mathbf{f}_j(ih,\bfxtildel_{j-1}(ih), \bfxtildel_j(ih),\bfxtildel_{j+1}(ih) )h,
\end{align}
with initial condition $\bfxtildel_j(0)=\bfxp_j$. This procedure is how do we obtain the step
\[
\bfxl(t,c) = \Phil(\bfxp, t,\bfxo(t\leq T,c), \bfW^{(c)})
\]
in a-MwG Algorithm \ref{alg:acc_l_mwg}. Note that in a-MwG, $\bfW^{(c)}$ are  Brownian motion realizations fixed during the loops. So when implementing \eqref{eqn:EMloc}, one use fixed $\mathcal{N}(\mathbf{0}_{b\times1},\mathbf{I}_b)$ samples $W^{(c)}_{i,j}$ in place of $W_{i,j}$, instead of generating new independent samples for $W_{i,j}$. 

The reason we need to do the copying step \eqref{eqn:EMbc} before the local Euler--Maruyama, is because the values of $\bfxl_{i_\star\pm (L+1)}(ih)$ are needed in the computation of $\bfxl_{i_\star\pm L}(ih)$ in \eqref{eqn:EMloc}. This means in real implementation, \eqref{eqn:EMbc} is only needed to be executed for the boundary blocks $j=i_\star\pm (L+1)$. This can save additional computation time and storage in practice. We do not choose to formulate \eqref{eqn:EMbc} and \eqref{eqn:EMloc} in this fashion for simplifying the notations.

Because \eqref{eqn:EMloc} only requires execution for $j\in \Bistar$, so the total cost of obtaining  $\bfxtildel$ is only $O(1)$, which is one order cheaper than obtaining $\bfxtildep$ directly with Euler--Maruyama. This partial computation naturally introduces an error. But similar to Theorem \ref{thm_con}, this error can be controlled through the local domain radius $L$, according to the following theorem:

\begin{thm}\label{thm_eul}
	Under Assumption \ref{lip}, for any given positive constant $C_d$, $j=1,...,m$ and $i=0,1,...,T/h$, we have  
	\begin{align}\label{th2res1}
	\begin{split}
	& \E[\|\tilde{\mb{x}}^{\text{\emph{l}}}_j(ih) - \tilde{\mb{x}}_j^{\text{\emph{p}}}(ih)\|^2] \\
	&\leq C_2(\mathbf{f}, \bfsigma) e^{2C_1(\mathbf{f}, \bfsigma) (1+h)ih} e^{-C_d (L+1)}  \|\mb{x}^{\emph{o}}_{i_\star}-\mb{x}^{\emph{p}}_{i_\star}\|^2,
	\end{split}
	\end{align}
	with $C_1(\mathbf{f}, \bfsigma), C_2(\mathbf{f}, \bfsigma)$ defined in (\ref{C_1}) and (\ref{C_2}). For any given constant $\epsilon$, if  the local domain radius $L$ satisfies
	\begin{align}\label{thm2_res3}
	L \geq \dfrac{\log{\Big( \dfrac{\epsilon}{C_2(\mathbf{f}, \bfsigma) \|\mathbf{x}^{\emph{o}}_{i_{\star}} - \mathbf{x}^{\emph{p}}_{i_{\star}}\|^2}   \Big)}}{-C_d}+\dfrac{2C_1(\mathbf{f}, \bfsigma) (1+h)}{C_d}T,
	\end{align}
	then $\E[\|\tilde{\mb{x}}^{\text{\emph{l}}}_j(ih) - \tilde{\mb{x}}_j^{\text{\emph{p}}}(ih)\|^2]\leq \epsilon$. 
\end{thm}
Note that as $h\to 0$, \eqref{th2res1} and \eqref{thm2_res3} converge to corresponding theoretical ones in Theorem \ref{thm_con}. Also note that based on \eqref{thm2_res3}, the choice of $L$ is independent of the  dimension $n$. 

\subsection{Local computation with Runge--Kutta}
When stochastic forcing is zero, our model \eqref{sys:original} is an ODE. This means we can use \black{$S=1$} in Algorithm \ref{alg:acc_l_mwg}, and it is not necessary to sample the Brownian motion. 4th order Runge--Kutta (RK4)  is the standard numerical method for ODE computation. 

When \black{applying} RK4 to the full model for obtaining $\bfxo(t\leq T)$ as a start, one runs the following 
\begin{align*}
\bfxtildeo_j((i+1)h) &= \bfxtildeo_j(ih) + \\
&\quad \dfrac{1}{6}(\mathbf{k}_j^{1,\text{o}}(i) + 2\mathbf{k}_j^{2,\text{o}}(i) + 2\mathbf{k}_j^{3,\text{o}}(i) + \mathbf{k}_j^{4,\text{o}}(i)),
\end{align*}
with
\begin{align}
\label{eqn:RK4}
\begin{split}
\mathbf{k}_j^{1,\text{o}}(i)   &= h \mathbf{f}_j(ih,   \bfxtildeo_{j-1}(ih) ,\bfxtildeo_{j}(ih),\bfxtildeo_{j+1}(ih))\\
\mathbf{k}_j^{2,\text{o}}(i) &= h \mathbf{f}_j(ih+\tfrac{h}{2},  \bfxtildeo_{j-1}(ih)+ \tfrac{\mathbf{k}_{j-1}^{1,\text{o}}(i)}{2}, \\
&\qquad \quad\bfxtildeo_{j}(ih) + \tfrac{\mathbf{k}_j^{1,\text{o}}(i)}{2}, \bfxtildeo_{j+1}(ih) + \tfrac{\mathbf{k}_{j+1}^{1,\text{o}}(i)}{2})\\
\mathbf{k}_j^{3,\text{o}}(i)& = h \mathbf{f}_j(ih+\tfrac{h}{2},  \bfxtildeo_{j-1}(ih) + \tfrac{\mathbf{k}_{\black{j-1}}^{2,\text{o}}(i)}{2} , \\
&\qquad \quad \bfxtildeo_{j}(ih)+ \tfrac{\mathbf{k}_j^{2,\text{o}}(i)}{2}, \bfxtildeo_{j+1}(ih) +\tfrac{\mathbf{k}_{j+1}^{2,\text{o}}(i)}{2})\\
\mathbf{k}_j^{4,\text{o}}(i) & = h \mathbf{f}_j(ih+h,  \bfxtildeo_{j-1}(ih)+ \mathbf{k}_{j-1}^{3,\text{o}}(i), \\
& \qquad \quad \bfxtildeo_{ \black{j}}(ih)  +\mathbf{k}_j^{3,\text{o}}(i), \bfxtildeo_{j+1}(ih)+ \mathbf{k}_{j+1}^{3,\text{o}}(i)).
\end{split}
\end{align}
This step is repeated for all $j=1,\ldots,m$. 

When applying RK4 to compute the local surrogate $\bfxl(t) = \Phil(\bfxp, t,\bfxo(t\leq T))$, we need not only the numerical approximation $\bfxtildeo(t)$ of $\bfxo(t)$, 
but also the intermediate values $\mathbf{k}^{s, \text{o} }_j(i)$ for $s=1,2,3,4$. These values will be directly taken as $\bfxtildel_j$ and its RK4 intermediate values for $j\in \Bistar^c$:
\begin{equation}
\label{eqn:RK4bc}
\bfxtildel_j(ih) = \bfxtildeo_j(ih), \quad \mathbf{k}^{s,\text{l}}_j(ih) = \mathbf{k}^{s,\text{o}}_j(ih). 
\end{equation}
New RK4 computation is only needed for blocks inside the local domain, and $\bfxtildel_j(ih)$ is obtained through iterating the following step for $j\in \Bistar$ and $i=1,\ldots, T/h$:
\begin{align}
\label{eqn:RK4b}
\begin{split}
&\bfxtildel_j((i+1)h) = \bfxtildel_j(ih)  + \dfrac{1}{6}(\mathbf{k}_j^{1,\text{l}}(i) + 2\mathbf{k}_j^{2,\text{l}}(i)\\
&\qquad \qquad \qquad \quad  + 2\mathbf{k}_j^{3,\text{l}}(i) + \mathbf{k}_j^{4,\text{l}}(i)).
\end{split}
\end{align}
Here the intermediate values $\mathbf{k}_j^{s,\text{l}}$ are defined in the same way as $\mathbf{k}_j^{s,\text{o}}$ in \eqref{eqn:RK4}, that is every instance of $\bfxtildeo_j$ is replaced by $\bfxtildel_j$, and every instance of $\mathbf{k}_j^{s,\text{o}}$ is replaced by $\mathbf{k}_j^{s,\text{l}}$. 
\black{Similar} as the local computation with Euler--Maruyama method, we note that \eqref{eqn:RK4bc} in reality is only needed for  \red{$\{j:L+1\leq |j-i_\star|\leq L+4\} $. To see this, note that to obtain $\bfxtildel_j((i+1)h)$ we need $\mathbf{k}^{4,l}_j$ with $j\in B_{i_\star}$. This needs $\mathbf{k}^{3,l}_j$ with $j\in \{j:|j-i_\star|\leq L+1\}$. Likewise, we need $\mathbf{k}^{4-s,l}_j$ with $j\in \{j:|j-i_\star|\leq L+s\}$ for $s=2,1,0$, where we let $\mathbf{k}^{0,l}_j=\bfxtildel_j(ih)$. We can exploit this to save both computation time and storage.}

%
%

By completing the procedures described above, we obtain the local surrogate $\bfxl(t) = \Phil(\bfxp, t,\bfxo(t\leq T))$. 
If $\bfxp$ is accepted in the MH step, it will be used as $\bfxo$ in the next iteration.
Since \eqref{eqn:RK4b} is repeated only for blocks within the local domain, the overall cost is $O(1)$.

While an approximation error analysis like Theorem \ref{thm_eul} in principle exists, we do not provide the detailed proof. This is partly because RK4 is much more complicated than Euler--Maruyama. Moreover, being an 4th order accuracy method, RK4 solution is very close to the exact solution of \eqref{sys:local}. So the approximation error can be learned directly from Theorem \ref{thm_con}. 

\subsection{Parallelization}
The a-MwG algorithm we proposed can be easily adapted to parallelization for shorter wall-clock computation time. 

First of all, in Algorithm \ref{alg:acc_l_mwg}, 
every repetition among all \black{$S$} Brownian motion realizations can be computed in parallel.
This is the case because we assume $\bfW^{(c)}$ are independent with each other.

Next, recall that in the local surrogate model (\ref{sys:local}), $\bfxl_j(t)$ needs renewed computation only for $j\in \Bistar$, which depends on $\bfxo_j(t)$ only at the boundary blocks $j=i_\star\pm (L+1)$.  Now consider another block index $j_\star$ such that $d(i_\star, j_\star)\geq 2L+2$. Then its local domain $B_{j_\star}$ and boundary blocks $j=j_\star\pm(L+1)$ have no intersection with the ones of $i_\star$. This means that if there are two Gibbs proposals $\bfx^{\text{p},i_\star}$ and $\bfx^{\text{p},j_\star}$, and they are different from the current iterate $\bfxo$ at the $i_\star$-th block and \black{ the} $j_\star$-th block respectively, then the computation of 
$\Phil(\bfx^{\text{p},i_\star}, t,\bfxo(t\leq T,c), \bfW^{(c)})$ and $\Phil(\bfx^{\text{p},j_\star}, t,\bfxo(t\leq T,c), \bfW^{(c)})$ will be independent. Thus we can compute them in parallel. 

In other words, when implementing a-MwG, at each Gibbs iteration, we can draw several proposals in parallel. Each proposal is different from the current iterate $\bfxo$ at one block index, and each pair of these block indices are of distant $2L+2$ or more apart. Then the local surrogate model (\ref{sys:local}) can be applied to each proposal in parallel, and \black{ the MH step} is run independently on each block. This will have no doubt to save \black{wall-clock} computation time. 

Recall that our SDE model \eqref{sys:original} can find its origin in stochastic partial differential equation (PDE)  such as the reaction-diffusion equation. 
With this in mind, our local computation scheme is closely connected to the PDE parallel computation scheme called ``domain decomposition". 
Both methods partition the computation domain into smaller local domains and try to do the computation only in the local domains. The difference is that the domain decomposition is mostly applied to linear PDEs, and its execution relies on manipulation of linear algebra. The local computation scheme introduced here is for inverse problems involving SDE formulation, which can be nonlinear. 

\section{Experiments}\label{sec:exp}

\subsection{Lorenz 96 model}
The Lorenz 96 model was introduced in \cite{L1995} as a simplified description for equatorial oceanic flows. It is commonly used in the testing of high dimensional data assimilation methods (see e.g. \cite{OHSZ2004,FHH2007,LSSS2016,ABN2016}). The component $x_{j}(t)$, for $j=1,...,n$, is governed by the following ODE
\begin{align}\label{lorenz}
\begin{split}
&\dfrac{dx_j(t)}{dt} = -x_{j-2}(t)x_{j-1}(t) + \\
&\qquad \qquad  x_{j-1}(t)x_{j+1}(t)-x_{j}(t) + 8,\ t \in [0,T],
\end{split}
\end{align}
where we let $ x_{-1}(t) = x_{n-1}(t)$, $ x_0(t)=x_n(t)$, $x_{n+1}(t) = x_1(t)$. The solution of \black{the Lorenz 96} model has an equilibrium distribution, which can be obtained by longtime simulation. We use a Gaussian approximation of it as the prior distribution, of which the mean vector and covariance matrix are obtained by the localization procedure described in \cite{MTM2019}.

We consider solving the inverse problem formulated as (\ref{inv}), where the underlying model (\ref{sys:original}) is given by (\ref{lorenz}) with $T=0.4$. It can be verified that the spatial interaction of this model is local in the form of (\ref{sys:original}) with a minimum choice of block size $b=2$. We consider that the observation is obtained with every other component, so $\mathbf{H}$ is an $n/2 \times n$ matrix with $\mathbf{H}_{i,2i-1} = 1$ for $i=1,...,n/2$ while other entries are 0. For simplicity, we only consider even $n$. For the noise term $\xi$, we assume  $ \xi \sim \mathcal{N}(\mathbf{0},\mathbf{I}_{n/2})$. The MwG sampler and a-MwG sampler are applied to draw $K$ posterior samples $\{\mathbf{x}^{k}: k= 1,...,K\}$ respectively. We implement the experiments for \eqref{lorenz} of  dimensions $n=40$ and $n=400$. We use RK4 with time step $h=0.01$ for both MwG and a-MwG. 

The choice of block size $b$ and local radius $L$ play important roles in our a-MwG sampling algorithm. The length of $b$ should be larger than the bandwidth of the prior covariance. This is discussed in detail in \cite{MTM2019}. As for $L$, from Theorems \ref{thm_con}-\ref{thm_eul} we see a smaller value leads to \black{a} larger error of a-MwG sampler but faster computation. While Theorems \ref{thm_con}-\ref{thm_eul} give theoretical lower bounds of $L$ to control the error, they might be pessimistic in practice. A better way is to simulate the proposal and rejection process within a-MwG, compare it with MwG, and choose $L$ so that the proposal acceptance rate and the ODE state at $T$ are close for two samplers. 

In specific, with a set of fixed parameters $n$, $b$ and $L$, we generate one $\bfxo$ from the prior distribution as in Algorithm \ref{alg:acc_l_mwg}. We replace the $j$-th block (we fix $j=1$ for simplicity)  of $\bfxo$ by the proposal $\bfxp_j$ and obtain  $\bfxp$. Recall in MwG, the ODE is computed through $\bfxp(t\leq T) = \Phi(\bfxp, t\leq T)$, and the acceptance probability $\alpha$ is given by \eqref{eqn:MH}. In a-MwG, we approximate the ODE solution with $\bfxl(T) =\Phil(\bfxp, t,\bfxo(t\leq T))$ and acceptance rate $\alpha' := \min\{ 1, \hat{p}^{\text{p}}_{l}/\hat{p}^{\text{o}}_{l}\}$  in Algorithm   \ref{alg:acc_l_mwg}. Based on these quantities, the following errors can serve as possible standards for choosing a proper $L$
\begin{align*}
\text{Err-}\Phi =\dfrac{\E_{\bfxo\sim p_0}  \|\bfxl(T) - \bfxp(T)\|_{\infty} }{ \E_{\bfxo\sim p_0}  \|\bfxp(T)\|_{\infty} },
\end{align*}
and
\begin{align*}
\text{Err-}\alpha = \E_{\bfxo\sim p_0} |\alpha-\alpha'|.
\end{align*}
We note that $\text{Err-}\Phi$ is the relative error between $\bfxl(T)$ and $\bfxp(T)$, while $\text{Err-}\alpha$ depicts their influence on acceptance rate. We use 500 independent samples $\bfxo$ from $p_0$ to approximate the average, and list the average errors in Table \ref{selectL}. From the table, we see that a choice of $L=4$ ($L=2$) for $b=2$ ($b=4$) is enough to keep the relative errors under $3 \%$, both in the senses of acceptance rate $\alpha$ and solution $\Phi$. Also note that all the error terms decrease as the length of $L$ increases. This verifies our previous theoretical analyses. 

\begin{table}[!htbp]
	\centering
	\caption{Err-$\alpha$ (with $\alpha$ varies within $[0.4846,0.5503]$) and Err-$\Phi$ for selecting a proper $L$, under model ($\ref{lorenz}$).}
	\label{selectL}
	\setlength{\tabcolsep}{0.5mm}{
		\begin{tabular}{|c|c|c|c|c|c|}
			\hline
			\multicolumn{2}{|c|}{} & \multicolumn{2}{|c|}{n=40} & \multicolumn{2}{|c|}{ n=400 } \\
			\hline
			$b$& $L$ &Err-$\alpha$&Err-$\Phi$& Err-$\alpha$&Err-$\Phi$\\
			\hline
			\multirow{3}*{2}&1&0.1858&0.3361&0.1531 &0.2940 \\
			\cline{2-6}
			~&2&0.0819&0.1855&0.0857&0.1486 \\
			\cline{2-6}
			~&4&0.0134&0.0235&0.0043&0.0194 \\
			\hline
			\multirow{2}*{4} &1&0.0817&0.1969 & 0.0548 & 0.1599\\
			\cline{2-6}
			&2&0.0119&0.0297& 0.0136&0.0209\\		
			\hline
	\end{tabular}}
\end{table}

Despite we have a criterion above to pick $L$, in below, we still test other choices of  $L$ for the purpose of comparison. After implementing the MwG and a-MwG sampling algorithms, the quantities of mean squared error (MSE),  mean sample variance (MSV) are calculated from the posterior samples generated
\begin{align*}
\text{MSE} = \dfrac{1}{n} \sum_{j=1}^{n} \|\bar{\mathbf{x}}_j - \mathbf{x}_j\|^2, 
\end{align*}
and 
\begin{align*}
\text{MSV} = \dfrac{1}{n(K-k_0)} \sum_{k=k_0+1}^{K} \sum_{j=1}^{n} \|\mathbf{x}^{ k}_{j} - \bar{\mathbf{x}}_j\|^2,
\end{align*}
where
\begin{align*}
\bar{\mathbf{x}} = \dfrac{1}{K-k_0} \sum_{k = k_0 +1}^{K} \mathbf{x}^{ k}.
\end{align*}
In above, the first $k_0$ samples are considered as burn-in and thrown away. Besides, the average acceptance rate (AR) and the total amount of computation time (CT) for generating the samples are also collected. The CT is calculated using a 2018 MacBook Pro with 2.2GHz Intel Core i7 processor. The related paremeters are set as $K = 100000$, $k_0 = 10000$, with corresponding simulation results presented in Table \ref{tab_l96}. We see that our a-MwG algorithm samples the posterior distribution correctly, yielding MSE and MSV results similar to \black{the ones for} MwG, and the acceleration effect is significant. Its acceptance rate is also similar to the one of MwG sampler, in particular for larger $L$.  

\begin{table*}[!htbp]
	\centering
	\caption{The results of AR, CT, MSE, MSV for 100000 samples drawn from MwG($b$) and a-MwG($b$,$L$) samplers, where $b, L$ are block size and radius length.}
	\label{tab_l96}
	\begin{tabular}{|c|c|c|c|c|c|c|c|c|}
		\hline
		& \multicolumn{4}{|c|}{n=40 } & \multicolumn{4}{|c|}{n=400 } \\
		\hline
		&AR&CT&MSE & MSV& AR& CT& MSE & MSV\\
		\hline
		MwG(2)&0.1343&2080&8.6902 &9.0255 &0.1149 &27334 &9.4972 & 10.2152 \\
		\hline 
		a-MwG(2,1)&0.1254&1440&9.5906&9.7888&0.1014 &12798 & 10.4869&12.5727 \\
		\hline
		a-MwG(2,2)&0.1310&1444&8.9625 &8.3956& 0.1090&12830 &9.8387 & 10.5445 \\
		\hline
		a-MwG(2,4)&0.1347&1492&9.2899&8.1608 &0.1152 &13194 &9.5599 &9.8135 \\
		\hline
		MwG(4)&0.0389&1041 &9.1626&8.2137 &0.0273 &13593 &10.4764 &9.5827\\
		\hline
		a-MwG(4,1)&0.0388&724&9.2161&8.3851 &0.0283 & 6460&9.6107 &9.9683\\
		\hline
		a-MwG(4,2)&0.0388&749&8.6808&8.8641 &0.0280 &6634 &9.9938&9.3718\\
		\hline			
	\end{tabular}
\end{table*}

\begin{figure*}[!htbp]
	\centering
	\caption{Shown in green are posterior samples generated from MwG(4) (top) and a-MwG(4,2) (bottom); Yellow ones are drawn from prior distribution; the true state is in red. }
	\includegraphics[width=16cm, height=5cm]{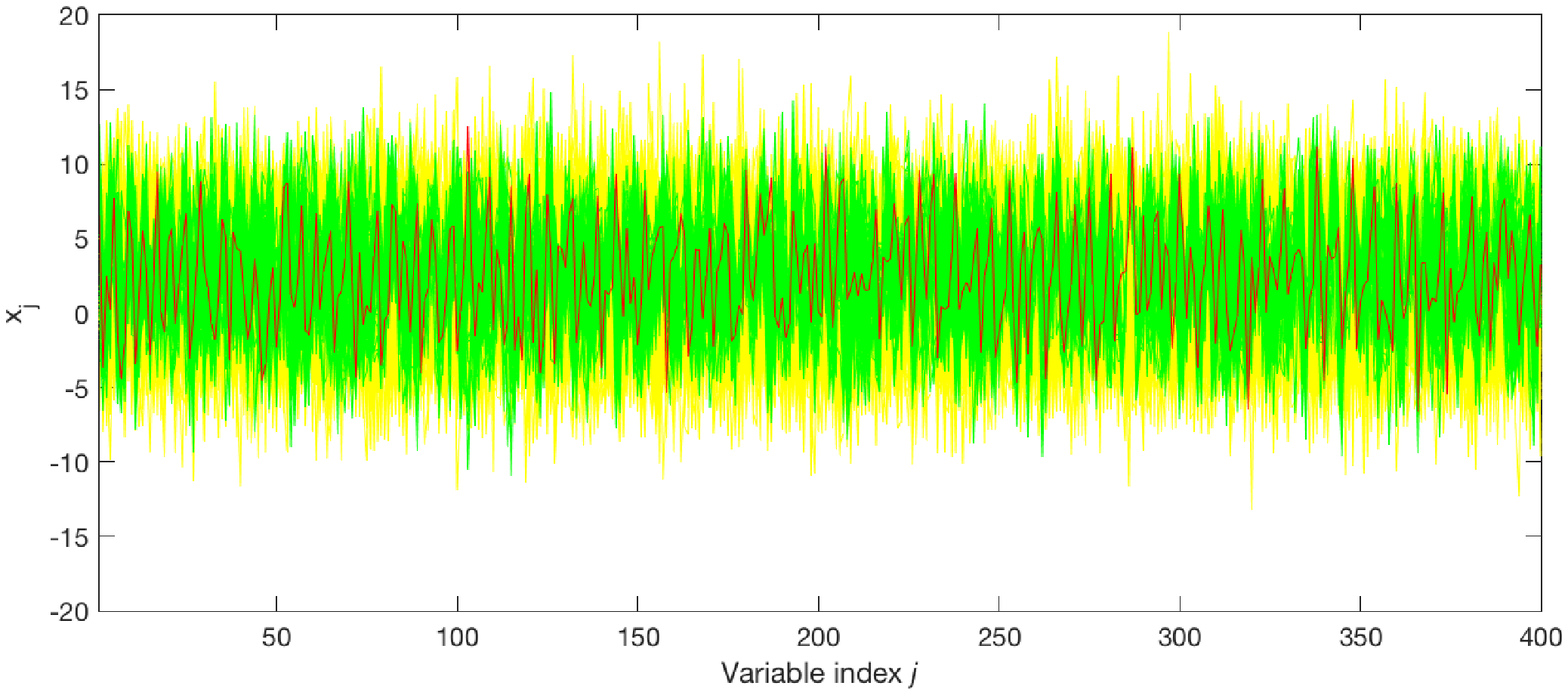}
	\includegraphics[width=16cm, height=5cm]{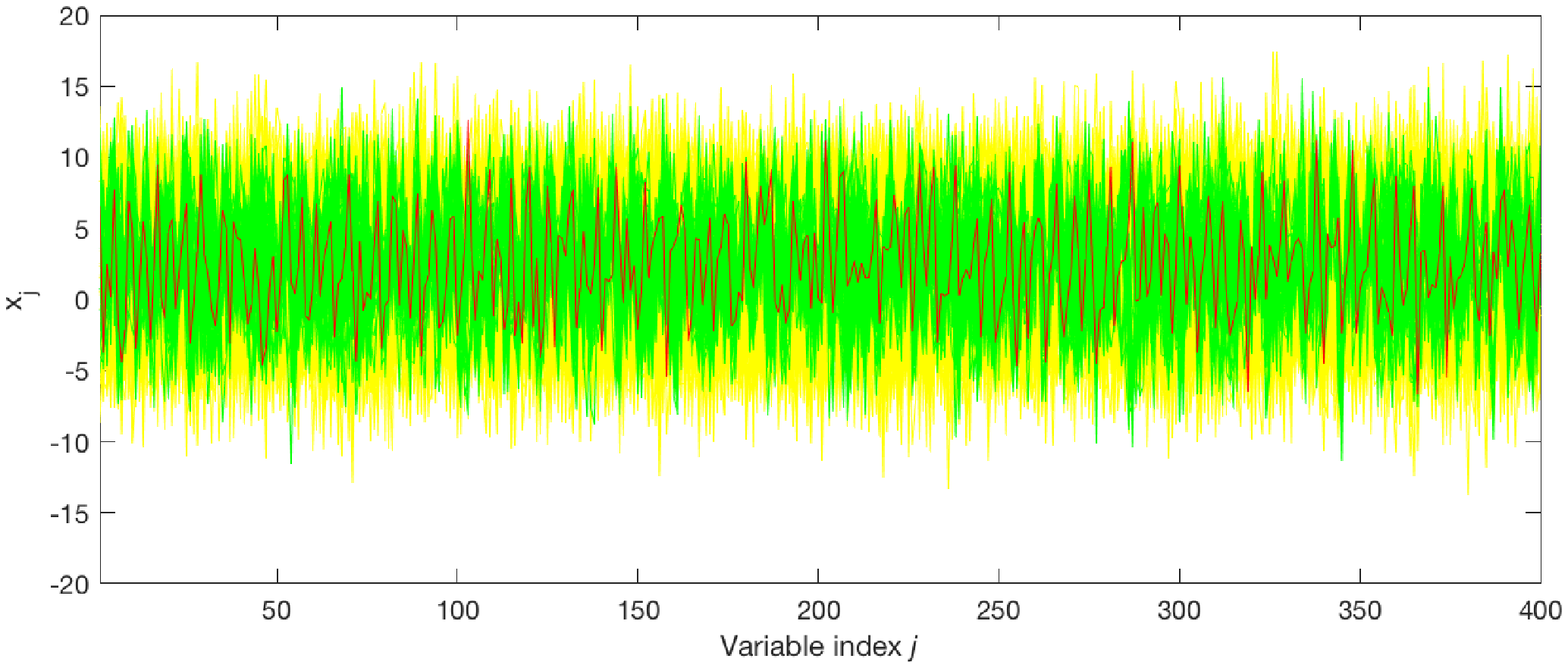}
	\label{fig_l96}
\end{figure*}

We also show  samples generated from two samplers MwG(4) and a-MwG(4,2) as a demonstration of the posterior distribution for the case $n=400$ in Figure \ref{fig_l96}. As a contrast, the same amount of prior samples and the true state are also shown. We see the two sampling algorithms perform similarly. 

\begin{figure}[!htbp]
	\centering
	\caption{Computation time of Lorenz 96 model when generating 1000 samples with different dimensions by MwG(4) (blue) and a-MwG(4,2) (green), for model (\ref{lorenz}).}
	\includegraphics[width=5.5cm,height= 6cm]{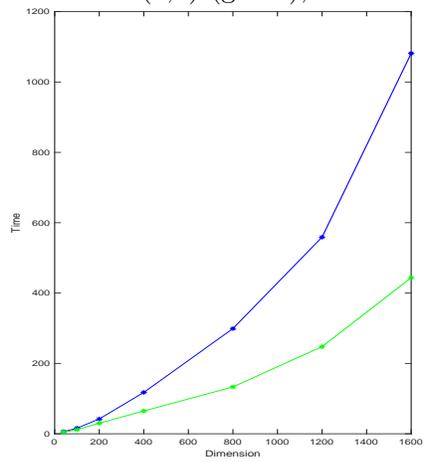}
	\label{time_l96}
\end{figure}

Specifically, for the computation time, we document its change when data dimension varies from 40 to 1600 in the examples of MwG(4) and a-MwG(4,2), in Figure \ref{time_l96}. It can be roughly seen that the computation cost of a-MwG increases linearly, compared with a quadratic increase for MwG. This verifies our previous theoretical analysis.

\subsection{Linearized stochastic flow model}
We study the following one-dimensional linearized stochastically forced dissipative advection equation in \cite{MH2012} (Section 6.3) and \cite{T2018}: 
\begin{align}\label{advection}
\begin{split}
&\dfrac{\partial f(x,t)}{\partial t} = w \dfrac{\partial f(x,t)}{\partial x} - \nu f(x,t) + \mu \dfrac{\partial^2f(x,t)}{\partial x^2} \\
& \qquad \qquad \  \ + \sigma_x\dfrac{\partial W(x,t)}{\partial t}, \ t \in [0,T],
\end{split}
\end{align}
where $W(x,t)$ is a white noise in both time and space. Applying the centered difference formula \black{with spatial grid size $l$ transforms} (\ref{advection}) into a time continuous linear stochastic system. The $n$-dimensional state $\mb x(t) = [x_1(t),..., x_n(t)]^{T}$ follows the SDE in below for $j = 1,...,n$, 
\begin{align}\label{sto}
\begin{split}
dx_j(t) &= (ax_{j-1}(t) +bx_{j}(t) + cx_{j+1}(t))dt \\
&\quad + \sigma_x dW_{j}(t),\\
x_0(t) &= x_n(t), \ x_{n+1}(t) = x_1(t), 
\end{split}
\end{align}
with 
\begin{align*}
a = \dfrac{\mu}{l^2} - \dfrac{w}{2l}, \ b = -\dfrac{2\mu}{l^2} - \nu, \ c = \dfrac{\mu}{l^2} + \dfrac{w}{2l},
\end{align*}
and $W_j(t)$ is a standard Brownian motion. We consider a regime with strong advection and weak damping by setting 
\begin{align*}
& l=0.2, \  \mu = 0.1,\ \nu = 0.1,  \ w=2, \ \sigma_x \equiv 0.1.
\end{align*}
We shall follow similar steps as in \black{the} last example to study the performance of our a-MwG sampling algorithm under this linear stochastic model. The only difference is the additional involvement of stochastic factors $\bfW^{(c)}$ for $c=1,...,\black{S}=100$, which are generated preliminarily. For simplicity, we  consider a prior distribution $p_0(\mathbf{x}) = \mathcal{N}(0, \mathbf{I}_n)$. We set $\xi \sim 0.1 \cdot \mathcal{N}(0, \mathbf{I}_n)$, $T=0.4$ and realize (\ref{sto}) by Euler--Maruyama with time step $h=0.01$. Other parameters remain the same as in the previous example, if not specified.

One computational issue rises when we evaluate the a-MwG acceptance probability. When the associated \black{dimension $n$ is} large, the sample likelihood $\hat{p}_l(\bfy|\bfx)$ is close to zero. So directly  evaluating the acceptance rate $\alpha = \min\{ 1, \hat{p}^{\text{p}}_{l}/\hat{p}^{\text{o}}_{l}\}$ in Algorithm \ref{alg:acc_l_mwg} may  leads to numerical singularity. This problem does not occur to the deterministic Lorenz 96 model, since $p_{l}(\bfy|\bfx)$ needs no approximation, and the difference of the log-likelihoods is explicitly accessible. 

To resolve this issue, we note the \black{pseudo-marginal} likelihood ratio is
\begin{align}\label{eqn:accept}
\dfrac{\hat{p}^{\text{p}}_{l}}{\hat{p}^{\text{o}}_{l}} = \dfrac{\sum_{c=1}^{\black{S}}\exp(-\tfrac12\|\bfy-\mathbf{H}\cdot \bfxl(T,c)\|^2_{\bfR})}{\sum_{c=1}^{\black{S}}\exp(-\tfrac12\|\bfy-\mathbf{H}\cdot \bfxo(T,c)\|^2_{\bfR})}.
\end{align}
Since $\bfxl(T,c)$ is different from $\bfxo(T,c)$ only for blocks inside the local domain $\Bistar$, the entries outside $\Bistar$ will provide no information for the sampling of block $\bfx_{i^*}$. Thus, dropping those entries will cause no significant difference for the evaluation of  (\ref{eqn:accept}). In other words, we approximate \eqref{eqn:accept} with a local version
\begin{align}\label{eqn:laccept}
\dfrac{\hat{p}^{\text{p}}_{l}}{\hat{p}^{\text{o}}_{l}}\approx \dfrac{\sum_{c=1}^{\black{S}}\exp(-\tfrac12\|\bfy_{\Bistar'}-\mathbf{H}_{i^*}\cdot \bfxl_{\Bistar}(T,c)\|^2_{\bfR_{i^*}})}{\sum_{c=1}^{\black{S}}\exp(-\tfrac12\|\bfy_{\Bistar'}-\mathbf{H}_{i^*}\cdot \bfxo_{\Bistar}(T,c)\|^2_{\bfR_{i^*}})}.
\end{align}
Here $\Bistar'=\{i: 2i-1\in \Bistar\}$ are observed indices in $\Bistar$,  $\mathbf{H}_{i^*}$ consists the $\Bistar'$ rows and $\Bistar$ columns of $\mathbf{H}$, and $\bfR_{i^*}$ consists of the $\Bistar'$ rows and columns of $\bfR$.  Formulation \eqref{eqn:laccept} avoids the computational singularity issue that may exists in (\ref{eqn:accept}), since the associated variables are no longer high dimensional. 

\black{Note that using \eqref{eqn:laccept} is equivalent to localizing the observation matrix as in \cite{MTM2019} Section 2.2. When \eqref{sys:original} is a linear SDE, $\bfy$ is a linear noisy observation of $\bfx(T,c)$, and the bias caused by such localization is discussed in \cite{MTM2019} Section 2.3. We expect similar bias control is also possible for general SDEs, while the exact verification is beyond the scope of this paper.  }
Also note that the local domain $\Bistar$ in (\ref{eqn:laccept}) does not need to be the same as the one used for SDE computation. For this test example,  we use 20 entries centered around the perturbed block.

\begin{figure*}[!htbp]
	\centering
	\caption{Shown in green are posterior samples generated from MwG(4) (top) and a-MwG(4,3) (bottom); Yellow ones are drawn from prior distribution; the true state is in red. }
	\label{sam_sto}
	\includegraphics[width=13cm,height=4.5cm]{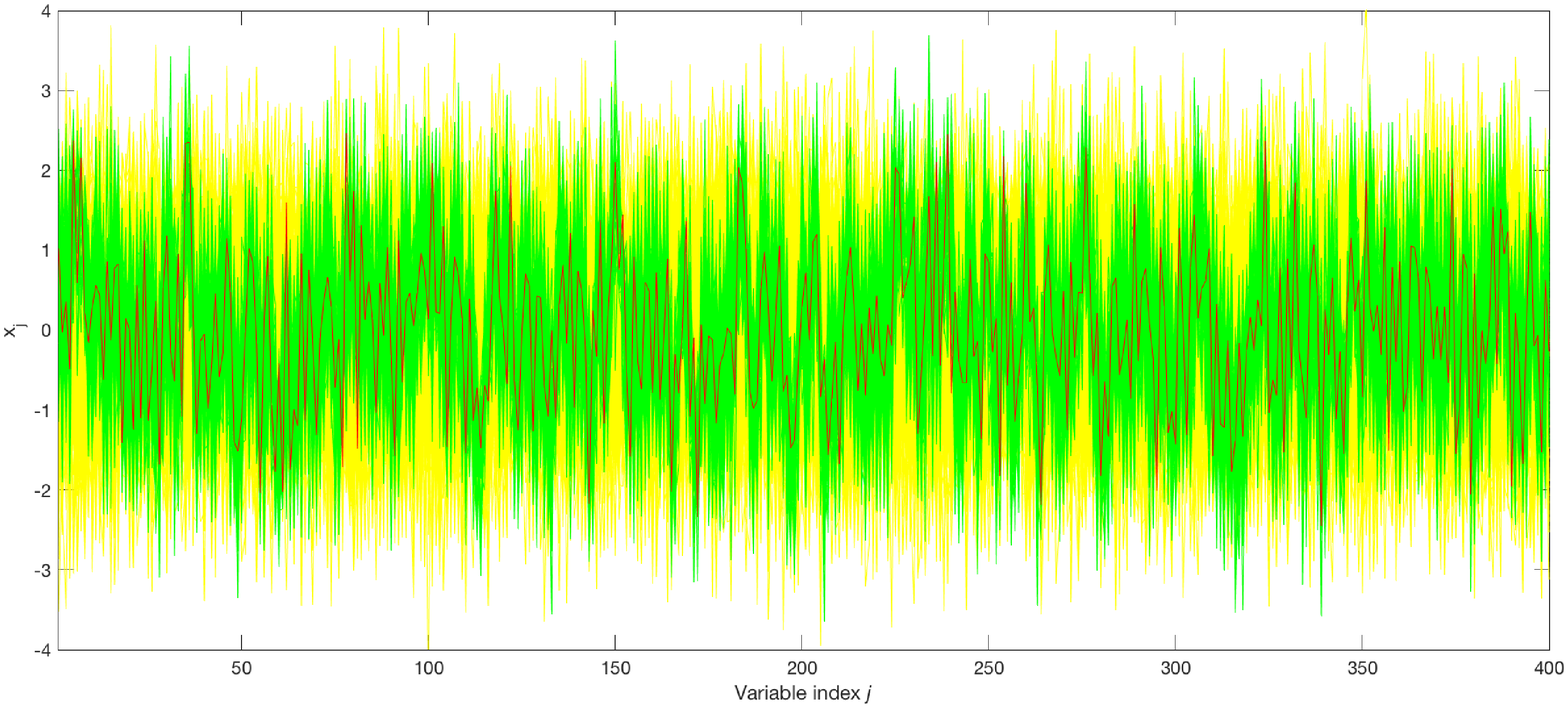}
	\includegraphics[width=13cm, height=4.5cm]{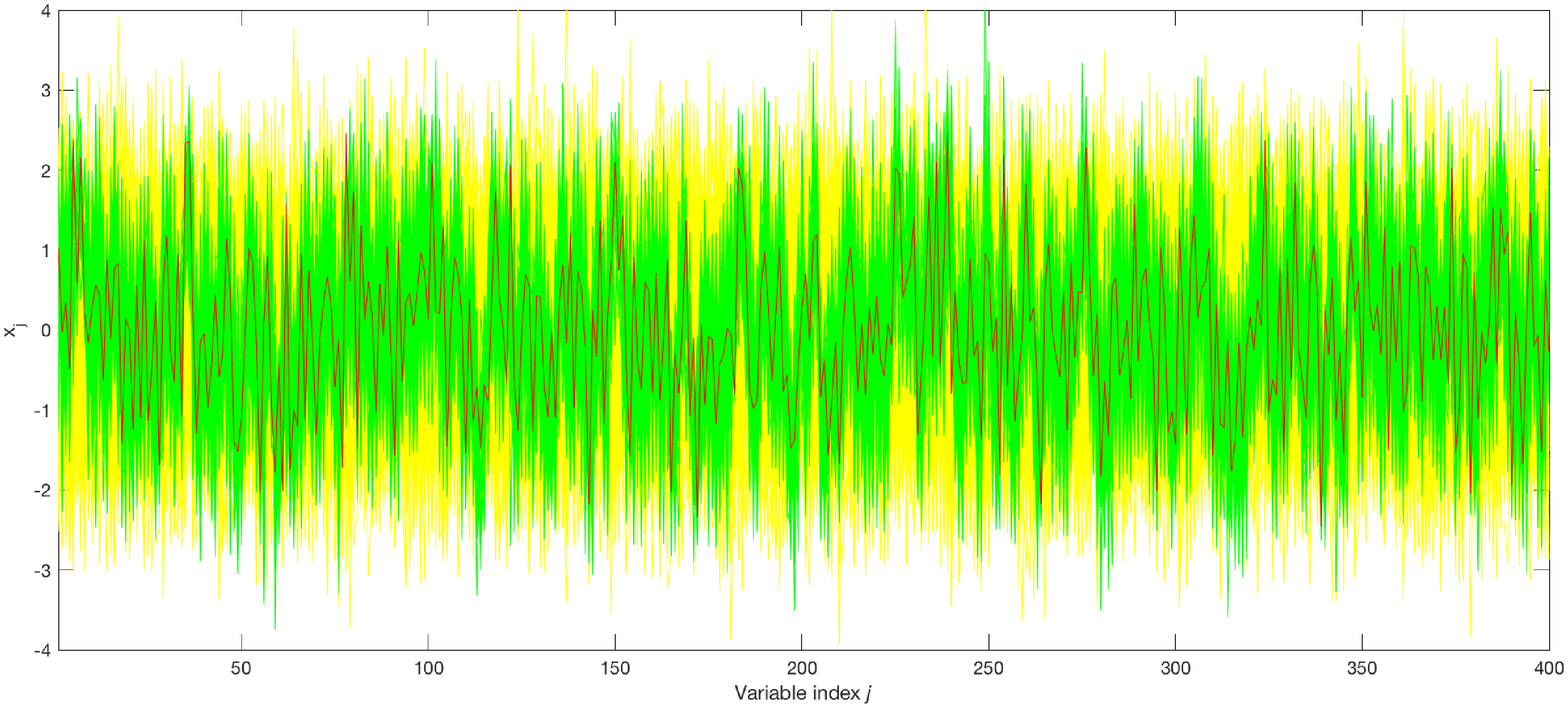}
\end{figure*}

\begin{figure}[!htbp]
	\centering
	\caption{Computation time of stochastic linear flow model when generating 100 samples with different dimensions by MwG(4) (blue) and a-MwG(4,2) (green), for model (\ref{sto}).}
	\includegraphics[width=5.5cm, height= 6cm]{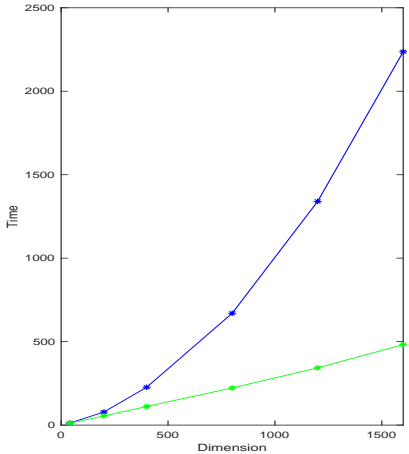}
	\label{time_sto}
\end{figure}

Before applying a-MwG sampler to solve the inverse problem, we use the same procedure described in the last example to choose the parameter $L$ of proper scale. The only difference is that all quantities depend on realizations of Brownian motions, and it is necessary to average over $\black{S}=100$ simulations.  We see from Table \ref{selectL_sto} that a minimum choice of $L=6$ ($L=3$) for $b=2$ ($b=4$) can even control the relative errors under only 1 percent. Finite sample performances of posterior distribution by using \black{MwG and a-MwG samplers} are presented in Table \ref{tab_sto} and Figure \ref{sam_sto}, from which we can see our sampling algorithm also works well for stochastic setting. \black{Figure \ref{time_sto} shows a more evident} linear relationship between the data dimension and the computation time for a-MwG, while a quadratic one for MwG. We note this is because for each proposal, the realization (\ref{sto}) is repeated for $\black{S}=100$ standard Brownian motion paths $\mathbf{W}$, dominating the total computation of the sampling algorithms and making the relationships more clearer.

\begin{table}[!htbp]
	\centering
	\caption{Err-$\alpha$ (with $\alpha$ varies within $[0.5045,0.5377]$) and Err-$\Phi$ for selecting a proper $L$, under model (\ref{sto}).}
	\label{selectL_sto}
	\setlength{\tabcolsep}{0.5mm}{
		\begin{tabular}{|c|c|c|c|c|c|}
			\hline
			\multicolumn{2}{|c|}{} & \multicolumn{2}{c}{n=40} & \multicolumn{2}{|c|}{ n=400 } \\
			\hline
			$b$& $L$ &Err-$\alpha$&Err-$\Phi$& Err-$\alpha$&Err-$\Phi$\\
			\hline
			\multirow{2}*{2}&2&0.0289&0.2200&0.0856 &0.1407 \\
			\cline{2-6}
			~&4&2.1e-04&0.0040&2.2e-04&0.0027 \\
			\hline
			\multirow{2}*{4} &1&0.0846&0.2189 & 0.0221 & 0.1604\\
			\cline{2-6}
			&2&1.3e-04&0.0043& 1.7e-04&0.0030\\		
			\hline
	\end{tabular}}
\end{table}

\begin{table*}[!htbp]
	\centering
	\caption{The results of AR, CT, MSE, MSV for 10000 samples draw from MwG($b$) and a-MwG($b$,$L$) samplers, where $b, L$ are block size and radius length. The exact posterior mean variance is listed in the end for comparison. }
	\label{tab_sto}
	\begin{tabular}{|c|c|c|c|c|c|c|c|c|}
		\hline
		& \multicolumn{4}{|c|}{n=40 } & \multicolumn{4}{|c|}{n=400 } \\
		\hline
		&AR&CT&MSE & MSV& AR& CT& MSE & MSV\\
		\hline
		MwG(2)&0.1342 &2418 &0.5953 & 0.5567&0.1494 &40617 &0.5799 &0.5413 \\
		\hline
		a-MwG(2,2)&0.1280&2052&0.6382&0.5708&0.1388 &20674 &0.6025 &0.5623 \\
		\hline
		a-MwG(2,4)&0.1344&2077&0.6009 &0.5302&0.1489 &21256 &0.5871 &0.5505  \\
		\hline
		MwG(4)&0.0367 &1193 &0.5854&0.5492 &0.0449 &21369 &0.6004 &0.5470\\
		\hline
		a-MwG(4,1)&0.0343&1007&0.5905&0.5244 &0.0423 &10511 &0.6077 &0.5417\\
		\hline
		a-MwG(4,2)&0.0353&1043&0.5903&0.5396 &0.0451 &10685 &0.5907 &0.5499\\
		\hline	
		Posterior & $\diagdown$&$\diagdown$&$\diagdown$& 0.5662 &$\diagdown$ & $\diagdown$ & $\diagdown$ & 0.5662\\
		\hline	
	\end{tabular}
\end{table*}

In fact, for this linear model, the posterior distribution of $\mb x$ given $\mb y $ can be derived explicitly. To derive it, we rewrite (\ref{sto}) in the following vertor form
\begin{align}\label{sto_vec}
d\mathbf{x}(t) = M\mathbf{x}(t)dt + \sigma d\mb W(t),
\end{align}
where $M$ is an $n$ by $n$ matrix with
\begin{align*}
&M_{1,1} = b, \ M_{1,2} = c, \ M_{1,n} = a,\\
&M_{j,j-1} = a, \ M_{j,j} = b, \ M_{j,j+1} = c,\text{for} \ j=2,...,n-1,\\
&M_{n,1} = c, \ M_{n,n} =b, \  M_{n,n-1} = a,
\end{align*}
and other entries are 0. According to Duhamel's formula, we have 
\begin{align*}
\mathbf{x}(T) = e^{MT} \mathbf{x}(0) + e^{MT}\int_{0}^{T}e^{-Mt}\sigma d\mb W(t).
\end{align*}
The observation model can therefore be written as
\begin{align}
\mathbf{y} = \mathbf{H} e^{MT} \mathbf{x}(0) + \mathbf{H} e^{MT}\int_{0}^{T}e^{-Mt}\sigma d\mb W(t) + \xi.
\end{align}
Recall that $ \mathbf{x}(0) \sim  \mathcal{N}(\mathbf{0}_{n \times 1},\Sigma_{\text{pri}})$ and $\xi \sim \mathcal{N}(\mathbf{0}_{n' \times 1},\bfR)$. It's known that for linear observation with  Gaussian prior distribution and noise, the posterior distribution is also of Gaussian type, and the posterior mean $\mathbf{m}_{\text{pos}}$ and covariance matrix $\Sigma_{\text{pos}}$ can be written as (\cite{MS1983}):
\begin{align*}
\Sigma_{\text{pos}}
&= (\Sigma_{\text{pri}}^{-1} + (\mathbf{H}e^{MT})^{T}(M'  + \bfR)^{-1}\mathbf{H}e^{MT})^{-1}, \\
\mathbf{m}_{\text{pos}}
&=  \Sigma_{\text{pos}} (\mathbf{H}e^{MT})^{T}( M'  + \bfR)^{-1}\mathbf{y},
\end{align*}
where 
\begin{align*}
M' = \mathbf{H} e^{MT}\cdot \int_{0}^{T} \sigma^2(e^{-Mt})^{T}e^{-Mt}dt\cdot (\mathbf{H} e^{MT})^{T}.
\end{align*}
The  mean variance of the true posterior distribution therefore can be computed explicitly with this formula. We document it in Table \ref{tab_sto} for comparison. We see that the MSVs based on the posterior samples draw from a-MwG are close to the exact values. These results show that our a-MwG can generate samples without much loss of accuracy while significantly reduce the computation time. 

\section{Conclusions}\label{sec:con}
In this paper, we discuss how to infer initial conditions of high dimensional SDEs with local interactions, when noisy observations are available. 
Previous research \cite{MTM2019} has shown that \black{the  Metropolis-within-Gibbs (MwG)} sampling has dimension independent convergence rate. 
However, each MwG iteration requires a computation cost of $O(n^2)$, with $n$ being the model dimension. 
The main contribution of this paper is that we introduce a reduced computation scheme to accelerate MwG implementation. 
We observe that in each updating step, the MwG proposal is only different from the original iterate at one parameter block, and the main difference caused also takes place near that block. 
Solving the SDEs within a local domain near this block, instead of doing the global computation for all entries, only involves $O(1)$ computation. 
Our accelerated algorithm, the a-MwG sampler, is proposed by integrating an approximate local computation into MwG sampling. The computation cost of each a-MwG iterate is  $O(n)$ in total. 
We derive rigorous bounds for the approximation errors and show that they can be controlled if  the local domain size is chosen to be larger than a constant threshold. 
We also discuss how to implement our local computation by using Euler--Maruyama and 4th order Runge--Kutta schemes. 
These methods are applied in our experimental studies of \black{the Lorenz 96} and linear stochastic flow model. Numerical results show that our sampling algorithm can be greatly accelerated without much loss of accuracy. 

\section*{acknowledgements}
	This work is supported by Singapore MOE-AcRF grant R-146-000-258-114 and  R-146-000-292-114.

\section*{Appendix}\label{sec:app}

\subsection*{Auxiliary lemmas and their proofs}\label{lemmas}

\begin{lem}\label{diag}
	An $n \times n$ matrix $A$ is circulant tridiagonal if $A_{j,i} = 0$ for all pairs of $(j,i)$ such that $d(j,i) > 1$ (Recall $d(j,i) = \min\{ |j-i|, |j+n-i|, |i+n-j| \}$).
	For such type of matrix, we have 
	\begin{align*}
	|(e^{A})_{j,i}| \leq e^{-C\cdot d(j,i)}\cdot e^{(e^{C}+e^{-C}+1)\rho}, \quad \forall \ j,i=1,2,...,n,
	\end{align*}
	where $\rho = \max\limits_{j,i = 1,2,...,n} |A_{j,i}|$, $C$ is any fixed positive constant.
\end{lem}
\begin{remark}
	If $A$ is further a tridiagonal matrix, we have
	\begin{align*}
	|(e^{A})_{j,i}| \leq e^{-C\cdot|j-i|}\cdot e^{(e^{C}+e^{-C}+1)\rho}, \quad \forall  \ j,i=1,2,...,n.
	\end{align*}
\end{remark}
$\textbf{Proof:}$ 
Firstly, by using mathematical induction, we prove that with any positive constant $C$, 
\begin{align}\label{Am_ji}
\begin{split}
&|(A^{m})_{j,i}| \\
& \leq (e^{C}+e^{-C}+1)^{m}\cdot\rho^m\cdot e^{-C\cdot d(j,i)}, \ \text{for} \ m=0,1,\cdots.
\end{split}
\end{align}
The conclusion is true for $m=0$ since $A^{0} = \mathbf{I}_n$ and $e^{-C\cdot d(j,i)} > 0$ with $e^{-C\cdot d(j,j)} = 1$. If the result is true for $m=k$, namely we have $|(A^{k})_{j,i}| \leq (e^{C}+e^{-C}+1)^{k}\cdot\rho^k\cdot e^{-C\cdot d(j,i)}$. Then, for $m=k+1$, we observe that
\begin{align}\label{A^k+1}
\begin{split}
&(A^{k+1})_{j,i} \\
& = (A^{k})_{j,i-1}A_{i-1,i} + (A^{k})_{j,i}A_{i,i} + (A^{k})_{j,i+1}A_{i+1,i},
\end{split}
\end{align}
with $(A^{k})_{j,0} = (A^{k})_{j,n}$, $(A^{k})_{j, n+1} = (A^{k})_{j,1}$, $A_{0,i} = A_{n,i}$, and $A_{n+1, i} = A_{1,i}$.
Since $|A_{j,i}| \leq \rho$, together with (\ref{Am_ji}) for $m=k$, (\ref{A^k+1}) gives us
\begin{align}\label{absA^k+1}
\begin{split}
&|(A^{k+1})_{j,i}| \\
& \leq | (A^{k})_{j,i-1}\|A_{i-1,i} | + |(A^{k})_{j,i}\|A_{i,i}| + |(A^{k})_{j,i+1}\|A_{i+1,i} | \\
& \leq (e^{C}+e^{-C}+1)^{k}\cdot\rho^k\cdot e^{-C\cdot d(j,i-1)} \cdot \rho \\
&\qquad + (e^{C}+e^{-C}+1)^{k}\cdot\rho^k\cdot e^{-C\cdot d(j,i)} \cdot \rho \\
& \qquad + (e^{C}+e^{-C}+1)^{k}\cdot\rho^k\cdot e^{-C\cdot d(j,i+1)} \cdot \rho.
\end{split}
\end{align}
We claim that 
\begin{align}\label{e^-C}
\begin{split}
&e^{-C\cdot d(j,i-1)} + e^{-C\cdot d(j,i)} + e^{-C\cdot d(j,i+1)} \\
&\leq (e^{C}+e^{-C}+1)e^{-C\cdot d(j,i)},
\end{split}
\end{align}
which will be proved later. Combining it with (\ref{absA^k+1}), we obtain
\begin{align}
|(A^{k+1})_{j,i}| \leq (e^{C}+e^{-C}+1)^{k+1}\cdot\rho^{k+1}\cdot e^{-C\cdot d(j,i)},
\end{align}
this finishes the proof of (\ref{Am_ji}). The claim (\ref{e^-C}) can be proved by the following case by case discussion. For the case $n=2k$ with $k\in \mathbb{N}^{+}$, we see $d(j,i) \leq k$ from the definition of $d(j,i)$. If $d(j,i) < k$, we have 
\begin{align*}
&e^{-C\cdot d(j,i-1)} + e^{-C\cdot d(j,i)} + e^{-C\cdot d(j,i+1)} \\
&= e^{-C\cdot (d(j,i)+1)} + e^{-C\cdot d(j,i)}  + e^{-C\cdot (d(j,i)-1)} \\
& = (e^{C}+e^{-C}+1)e^{-C\cdot d(j,i)}.
\end{align*}
Otherwise, if $d(j,i) = k$, we have 
\begin{align*}
&e^{-C\cdot d(j,i-1)} + e^{-C\cdot d(j,i)} + e^{-C\cdot d(j,i+1)} \\
&= e^{-C\cdot (d(j,i)-1)} + e^{-C\cdot d(j,i)}  + e^{-C\cdot (d(j,i)-1)} \\
& \leq (e^{-C}+e^{-C}+1)e^{-C\cdot d(j,i)}.
\end{align*}
For $n=2k-1$ with $k\in \mathbb{N}^{+}$, we observe $d(j,i) \leq k-1$. If $d(j,i) < k-1$,  we have 
\begin{align*}
&e^{-C\cdot d(j,i-1)} + e^{-C\cdot d(j,i)} + e^{-C\cdot d(j,i+1)} \\
&= e^{-C\cdot (d(j,i)-1)} + e^{-C\cdot d(j,i)}  + e^{-C\cdot (d(j,i)+1)} \\
&= (e^{C}+e^{-C}+1)e^{-C\cdot d(j,i)}.
\end{align*}
Otherwise, if $d(j,i) =k-1$, we have 
\begin{align*}
&e^{-C\cdot d(j,i-1)} + e^{-C\cdot d(j,i)} + e^{-C\cdot d(j,i+1)} \\
&= e^{-C\cdot (d(j,i)-1)} + e^{-C\cdot d(j,i)}  + e^{-C\cdot d(j,i)} \\
& \leq (e^{-C}+1+1)e^{-C\cdot d(j,i)}.
\end{align*}
For all the cases discussed above, we always have (\ref{e^-C}).

With (\ref{Am_ji}) proved, we have
\begin{align*}
& |(e^{A})_{j,i}| =  \Big| \sum_{m=0}^{\infty} \frac{1}{m!}\cdot(A^m)_{j,i} \Big| \leq \sum_{m=0}^{\infty} \frac{1}{m!}\cdot|(A^m)_{j,i}| \\
& \leq \sum_{m=0}^{\infty} \frac{1}{m!}\cdot(e^{C}+e^{-C}+1)^{m}\cdot\rho^m\cdot e^{-C\cdot d(j,i)} \\
&= e^{(e^{C}+e^{-C}+1)\rho} \cdot e^{-C\cdot d(j,i)}.
\end{align*}
The proof is complete. 
\hfill $\Box$

\begin{lem}[Gronwall's inequality] \label{gron}
	Given an $n \times n$ matrix $M$ whose entries are positive, and if the two $n \times 1$ vectors $\bfx(t)$, $\mathbf{b}(t)$ satisfy 
	\begin{align}\label{Mx}
	M\cdot \bfx(t) + \mathbf{b}(t) \succeq \dfrac{d\bfx(t)}{dt}, \ \text{for} \ t \in [0, T],
	\end{align}
	(Recall $\succeq$ is entriwise inequality, as defined in Section \ref{pre}), then
	\begin{align}\label{eMT}
	e^{MT}\cdot \bfx(0) + \int_0^T e^{M(T-t)} \cdot \textbf{b}(t) dt\succeq \bfx(T).
	\end{align}
\end{lem}
\textbf{Proof:} Consider an $n$-dimensional column vector $\bfy(t)$ defined as
\begin{align*}
\frac{d\bfy(t)}{dt} = M\cdot \bfy(t) + \mathbf{b}(t), \ \text{for} \ t \in [0, T],
\end{align*}
with initial value $\bfy(0) = \bfx(0) + \epsilon \cdot \mathbf{1}_{n \times 1}$, where $\epsilon$ is a positive constant. Its solution can be written as 
\begin{align*}
\bfy(T) = e^{MT}\cdot ( \bfx(0) + \epsilon \cdot \mathbf{1}_{n\times 1} ) + \int_0^T e^{M(T-t)} \cdot \textbf{b}(t) dt.
\end{align*}
Define $\mathbf{z}(t) = \bfy(t) - \bfx(t)$ for $t \in [0,T]$, then $\mathbf{z}(0) = \epsilon \cdot \mathbf{1}_{n \times 1} \succeq \mathbf{0}_{n \times 1}$. And according to (\ref{Mx}), we have
\begin{align*}
\frac{d\mathbf{z}(t)}{dt} &= \frac{d\bfy(t)}{dt} - \frac{d\bfx(t)}{dt} \\
& = M\cdot \bfy(t) + \mathbf{b}(t) - \frac{d\bfx(t)}{dt} \\
& \succeq M \cdot (\bfy(t) - \bfx(t)) = M \cdot \mathbf{z}(t).
\end{align*}
We define $\tau = \inf\{ t: \text{there exists at least one index} \ i,\\ \text{such} \ \text{that} \ \mathbf{z}_{i}(t) < 0. \}$, and note that $\tau > 0$ by continuity of $\mathbf{z}(t)$. According to the definition, we have $\mathbf{z}(t) \succeq \mathbf{0}_{n \times 1}$ for $t \in [0,\tau)$. Together with that all the entries of $M$ are positive, we have $\frac{d\mathbf{z}(t)}{dt} \succeq M \cdot \mathbf{z}(t) \succeq \mathbf{0}_{n \times 1}$ for $t \in [0, \tau)$. It implies that $\mathbf{z}(t)$ is entriwisely nondecreasing in $[0, \tau)$, thus $\mathbf{z}(t) \succeq \mathbf{z}(0) = \epsilon \cdot \mathbf{1}_{n \times 1}$ for $t \in [0, \tau)$, namely
\begin{align}\label{eMt}
\begin{split}
&e^{Mt}\cdot ( \bfx(0) + \epsilon \cdot \mathbf{1}_{n\times 1} ) + \int_0^t e^{M(t-s)} \cdot \textbf{b}(s) ds - \bfx(t) \\
& \succeq \epsilon \cdot \mathbf{1}_{n \times 1}.
\end{split}
\end{align}

We show that $\tau > T$, which can be proved by contradiction. Suppose that $\tau \leq T$. From the definition of $\tau$ and the previous analyses, we know that $\mathbf{z}(t) \succeq \epsilon \cdot \mathbf{1}_{n \times 1}$ when $t \in [0,\tau)$, and there exists at least one index, say $i$, such that $\mathbf{z}_{i}(\tau) < 0$. Thus we have $\lim\limits_{t \rightarrow \tau^{-}} \mathbf{z}_{i}(t) - \mathbf{z}_{i}(\tau) \geq \epsilon > 0$. This result contradicts with the continuity of $\mathbf{z}(t)$ for $t \in [0,T]$ and gives us the result $\tau > T$.

The conclusion (\ref{eMT}) is obtained by taking $\epsilon \rightarrow 0$ in (\ref{eMt}), together with $\tau > T$.
\hfill $\Box$

\begin{lem}\label{ineq}
	For two $n \times n$ matrices $M$ and $N$ whose entries are positive, if 
	\begin{align}\label{MN}
	M \preceq  N,
	\end{align}
	we have
	\begin{align}\label{e^M}
	e^{M} \preceq e^{N}.
	\end{align}
\end{lem}
\textbf{Proof:} By mathematical induction, we firstly prove 
\begin{align}\label{M^k}
M^{k} \preceq N^{k}, \ \text{for} \ k=0,...,\infty.
\end{align}
The result is true for $k=0$, since $M^{0} = N^{0} = \mathbf{I}_n$. If the result is true for $k=K$, namely we have
\begin{align}\label{MK}
M^{K} \preceq N^{K}.
\end{align}
Then for $k= K+1$, since
\begin{align*}
(M^{K+1})_{i,j} &= \sum_{s=1}^{n}(M^{K})_{i,s}M_{s,j}, \\ (N^{K+1})_{i,j} &= \sum_{s=1}^{n}(N^{K})_{i,s}N_{s,j}, \ \ \text{for} \ i,j = 1,...,n,
\end{align*}
the result $(M^{K+1})_{i,j} \leq (N^{K+1})_{i,j}$ follows from (\ref{MN}) and (\ref{MK}). 

According to the definition of matrix exponential, we have 
\begin{align*}
e^{M} - e^{N} = \sum_{k=0}^{+\infty} \dfrac{1}{k!} (M^{k} - N^{k}),
\end{align*}
the conclusion (\ref{e^M}) naturally follows from (\ref{M^k}). \hfill $\Box$

\begin{lem}[Volterra's inequality]\label{gron_s} 
	Given an $n \times n$ matrix $M$ whose entries are positive, if the two $n \times 1$ vectors $\bfx(t)$, $\mathbf{b}(t)$ satisfy
	\begin{align}\label{xt}
	\bfx(t) \preceq \mathbf{b}(t) + \int_{0}^{t}M\cdot \bfx(s) ds, \ \text{for} \ t \in [0, T],
	\end{align}
	we have 
	\begin{align}\label{xT}
	\bfx(T) \preceq \mathbf{b}(T) + \int_{0}^{T} e^{M(T-t)}\cdot M \cdot \mathbf{b}(t)dt.
	\end{align}
\end{lem}
\textbf{Proof:} We define $\bfy(t) = \int_{0}^{t} M\cdot \bfx(s)ds$ for $t \in [0, T]$, then (\ref{xt}) gives us 
\begin{align}\label{xtb}
\bfx(t) \preceq \mathbf{b}(t) + \bfy(t), \ \text{for} \ t \in [0,T].
\end{align}
And we observe that
\begin{align*}
\dfrac{d\bfy(t)}{dt} = M\cdot \bfx(t) \preceq M\cdot\mathbf{b}(t) + M\cdot\bfy(t),
\end{align*}
with $\bfy(0) = 0$. According to Lemma \ref{gron}, we have 
\begin{align}\label{yT}
\bfy(T) \preceq \int_{0}^{T} e^{M(T-t)}\cdot M \cdot \mathbf{b}(t)dt.
\end{align}
The conclusion (\ref{xT}) follows from (\ref{xtb}) with $t=T$ and (\ref{yT}). \hfill $\Box$

\subsection*{Main lemmas and their proofs}

\begin{lem}\label{bound_opz_l}
	Under Assumption \ref{lip}, with $\mb{x}^{\emph{o}}(t), \mb{x}^{\emph{p}}(t)$ for $t \in [0,T]$ defined in Section \ref{def_xop}, we have, for $j=1,...,m$, 
	\begin{align}\label{lExop2}
	\begin{split}
	&\E[\|\mb{x}^{\text{\emph{o}}}_j(t) - \mb{x}_j^{\text{\emph{p}}}(t)\|^2] \\
	&\leq \| \mb{x}^{\emph{o}}_{i_\star}-\mb{x}^{\emph{p}}_{i_\star} \|^2\cdot e^{C_1(\mathbf{f}, \bfsigma) t} \cdot e^{-C_d\cdot d(j,i_{\star})},
	\end{split}
	\end{align}
	where $C_d$ is any given positive constant, $C_1(\mathbf{f}, \bfsigma)$ is defined as (\ref{C_1}).
\end{lem}
\textbf{Proof:} Recall that $\mb{x}^{\text{o}}_j(t)$, $\mb{x}_j^{\text{p}}(t)$ are solutions of 
\begin{align*}
& d\bfx_j(t) = \mathbf{f}_j(t, \bfx_{j-1}(t), \bfx_j(t),\bfx_{j+1}(t) )dt \\
&\qquad \qquad + \bm{\sigma}_j(t,\bfx_j(t))d\bfW_j(t), \ \text{for} \ j=1,...,m,\\
& \bfx_{0}(t) = \bfx_{m}(t), \ \bfx_{m+1}(t) = \bfx_{1}(t), \  t \in [0,T],
\end{align*}
with corresponding initial values $\mb{x}^{\text{o}}_j(0)$, $\mb{x}_j^{\text{p}}(0)$, which only differ at the $i_{\star}$-th entry. Thus, we have
\begin{align*}
d(\bfx^{\text{o}}_j(t) - \bfx^{\text{p}}_j(t)) = \Delta_{\mathbf{f}_j}(t) dt + \Delta_{\bm\sigma_{j}}(t)d\bfW_j(t),
\end{align*} 
with 
\begin{align*}
& \Delta_{\mathbf{f}_j}(t)  = \mathbf{f}_j(t, \bfx^{\text{o}}_{j-1}(t), \bfx^{\text{o}}_j(t),\bfx^{\text{o}}_{j+1}(t) ) \\
& \qquad \qquad- \mathbf{f}_j(t, \bfx^{\text{p}}_{j-1}(t), \bfx^{\text{p}}_j(t),\bfx^{\text{p}}_{j+1}(t) ),\\
&  \Delta_{\bm\sigma_{j}}(t) = (\bm{\sigma}_j(t,\bfx^{\text{o}}_j(t))- \bm{\sigma}_j(t,\bfx^{\text{p}}_j(t))).
\end{align*}
According to It$\hat{\text{o}}$'s formula,
\begin{align}\label{dxop2}
\begin{split}
&d\|\bfx^{\text{o}}_j(t) - \bfx^{\text{p}}_j(t)\|^2\\
&= (2(\bfx^{\text{o}}_j(t) - \bfx^{\text{p}}_j(t))^{T}\Delta_{\mathbf{f}_j}(t) +   \|\Delta_{\bm\sigma_{j}}(t)\|^2)dt \\
& \quad +2(\bfx^{\text{o}}_j(t) - \bfx^{\text{p}}_j(t))^{T}\Delta_{\bm\sigma_{j}}(t) d\bfW_j(t).
\end{split}
\end{align} 
Its solution can be written as
\begin{align*}
&\|\bfx^{\text{o}}_j(t) - \bfx^{\text{p}}_j(t)\|^2 - \|\bfx^{\text{o}}_j(0) - \bfx^{\text{p}}_j(0)\|^2 \\
&= \int_{0}^{t}(2(\bfx^{\text{o}}_j(s) - \bfx^{\text{p}}_j(s))^{T}\Delta_{\mathbf{f}_j}(s) +   \|\Delta_{\bm\sigma_{j}}(s)\|^2)ds\\
& \quad + \int_{0}^{t}2(\bfx^{\text{o}}_j(s) - \bfx^{\text{p}}_j(s))^{T}\Delta_{\bm\sigma_{j}}(s) d\bfW_j(s).
\end{align*}
After taking expectation with respect to both sides of above formula, and according to Assumption \ref{lip} and Cauchy-Schwartz inequality, we have
\begin{align}\label{Exop2}
\begin{split}
&\E[\|\bfx^{\text{o}}_j(t) - \bfx^{\text{p}}_j(t)\|^2] - \|\bfx^{\text{o}}_j(0) - \bfx^{\text{p}}_j(0)\|^2 \\
&= \int_{0}^{t}(2(\bfx^{\text{o}}_j(s) - \bfx^{\text{p}}_j(s))^T\Delta_{\mathbf{f}_j}(s) +   \|\Delta_{\bm\sigma_{j}}(s)\|^2)ds\\
& \leq  \int_{0}^{t}\E[\|\bfx^{\text{o}}_j(s) - \bfx^{\text{p}}_j(s)\|^2]ds \\
& \quad +  \int_{0}^{t}\E[\|\Delta_{\mathbf{f}_j}(s)\|^2 + \|\Delta_{\bm\sigma_{j}}(s)\|^2]ds\\
& \leq  C_{\mathbf{f}}\int_{0}^{t}\E[\|\bfx^{\text{o}}_{j-1}(s) - \bfx^{\text{p}}_{j-1}(s)\|^2]ds \\
& \quad+ (C_{\mathbf{f}} +C_{\bm\sigma} + 1) \int_{0}^{t}\E[\|\bfx^{\text{o}}_j(s) - \bfx^{\text{p}}_j(s)\|^2]ds \\
& \quad+ C_{\mathbf{f}}  \int_{0}^{t}\E[\|\bfx^{\text{o}}_{j+1}(s) - \bfx^{\text{p}}_{j+1}(s)\|^2]ds. 
\end{split}
\end{align}
Equivalently, we can write (\ref{Exop2}) into the \black{following} vector form 
\begin{align*}
\vec{\Delta}(t) \preceq \int_{0}^{t}M\cdot \vec{\Delta}(s)ds + \vec{\Delta}(0),
\end{align*}
where $\vec{\Delta}(t)$ is an $m \times 1$ vector whose $j$-th entry is $\E[\|\mb{x}^{\text{o}}_j(t) - \mb{x}^{\text{p}}_j(t)\|^2] $; $M$ is an $m \times m$ matrix with 
\begin{align*}
&M_{1,1} =C_{\bfsigma} + C_{\mathbf{f}}+1,  M_{1,2} = M_{1,n} =C_{\mathbf{f}}, \\
&M_{j,j} =C_{\bfsigma}+ C_{\mathbf{f}}+1, M_{j,j+1} = M_{j,j-1} = C_{\mathbf{f}},\\
& \qquad \text{for} \ j=2,...,m-1,\\
& M_{m,m} =C_{\bfsigma}+ C_{\mathbf{f}}+1, M_{m,1} = M_{m,m-1} = C_{\mathbf{f}},
\end{align*}
and other entries are 0; moreover, we have $\vec{\Delta}_{i_{\star}}(0) = \| \mb{x}^{\text{o}}_{i_\star}-\mb{x}^{\text{p}}_{i_\star} \|^2$ and other entries of $\vec{\Delta}(0)$ are 0. According to Lemma \ref{gron_s}, we have
\begin{align*}
\vec{\Delta}(t) \preceq \vec{\Delta}(0) + \int_{0}^{t} e^{M(t-s)}\cdot M \cdot \vec{\Delta}(0)ds.
\end{align*}
We observe that for the $m$-dimensional column vector $(M \cdot \vec{\Delta}(0))$, its $i_{\star}-1, i_{\star}, i_{\star}+1$-th entries are $C_{\mathbf{f}}\| \mb{x}^{\text{o}}_{i_\star}-\mb{x}^{\text{p}}_{i_\star} \|^2, (C_{\mathbf{f}} + C_{\bfsigma}+1)\| \mb{x}^{\text{o}}_{i_\star}-\mb{x}^{\text{p}}_{i_\star} \|^2 , C_{\mathbf{f}}\| \mb{x}^{\text{o}}_{i_\star}-\mb{x}^{\text{p}}_{i_\star} \|^2 $ respectively, while other entries are 0. Together with Lemma \ref{diag} for $C=C_d$, $\max\limits_{j,i = 1,2,...,m} |M_{j,i}| = (C_{\mathbf{f}}+C_{\bfsigma}+1)$, (\ref{e^-C}) and the definition of $C(\mathbf{f}, \bfsigma)= (e^{C_d}+e^{-C_d}+1)(C_{\mathbf{f}}+C_{\bfsigma}+1)$, we have
\begin{align*}
&\vec{\Delta}_j(t) \\
&\leq  \vec{\Delta}_j(0) + \int_{0}^{t}|(e^{M(t-s)})_{j,i_{\star}-1}|\cdot (M \cdot \vec{\Delta}(0) )_{i_{\star}-1,1}ds  \\
&\quad  + \int_{0}^{t}|(e^{M(t-s)})_{j,i_{\star}}|\cdot (M \cdot \vec{\Delta}(0))_{i_{\star},1}ds \\
& \quad  + \int_{0}^{t}|(e^{M(t-s)})_{j,i_{\star}+1}|\cdot (M\cdot \vec{\Delta}(0))_{i_{\star}+1,1}ds \\
& \leq \vec{\Delta}_j(0)  +  (C_{\mathbf{f}}+ C_{\bfsigma}+1)\| \mb{x}^{\text{o}}_{i_\star}-\mb{x}^{\text{p}}_{i_\star} \|^2  \int_{0}^{t} e^{C_1(\mathbf{f}, \bfsigma) (t-s) }ds \\
& \quad \cdot  (e^{-C_d\cdot d(j,i_{\star}-1)} + e^{-C_d\cdot d(j,i_{\star})} + e^{-C_d\cdot d(j,i_{\star}+1)})  \\
& \leq \vec{\Delta}_j(0) + \| \mb{x}^{\text{o}}_{i_\star}-\mb{x}^{\text{p}}_{i_\star} \|^2 \cdot(e^{C_1(\mathbf{f}, \bfsigma)t }-1)  e^{-C_d\cdot d(j,i_{\star})} \\
& \leq \| \mb{x}^{\text{o}}_{i_\star}-\mb{x}^{\text{p}}_{i_\star} \|^2\cdot e^{C_1(\mathbf{f}, \bfsigma)\cdot t } \cdot e^{-C_d\cdot d(j,i_{\star})},
\end{align*}
where the last inequality is derived by considering two cases: when $j=i_{\star}$, $ \vec{\Delta}_j(0) =  \| \mb{x}^{\text{o}}_{i_\star}-\mb{x}^{\text{p}}_{i_\star} \|^2$ and $d(j,i_{\star}) = 0$, the result is true; when $j \neq i_{\star}$, $ \vec{\Delta}_j(0) = 0$, the result can also be verified. Recall that $\vec{\Delta}_j(t) = \E[\|\mb{x}^{\text{o}}_j(t) - \mb{x}^{\text{p}}_j(t)\|^2] $, the proof is complete.
\hfill $\Box$

\begin{lem}\label{bound_lpz}
	Under Assumption \ref{lip}, with $\mb{x}^{\emph{o}}(t)$, $\mb{x}^{\emph{l}}(t)$, $\mb{x}^{\emph{p}}(t)$ defined for $t \in [0,T]$ in Section \ref{def_xop} and \ref{def_xlp}, we have, for $j=1,...,m$, 
	\begin{align}\label{xlp2}
	\begin{split}
	&\E[ \|\bfx^{\text{\emph{l}}}_j(t) - \mb{x}_j^{\text{\emph{p}}}(t)\|^2 ]\\
	& \leq C_2(\mathbf{f}, \bfsigma) \| \mb{x}^{\emph{o}}_{i_\star}-\mb{x}^{\emph{p}}_{i_\star} \|^2 e^{2C_1(\mathbf{f}, \bfsigma) t} e^{-C_d\cdot(L+1)},
	\end{split}
	\end{align} 
	where $C_d$ is any given positive constant, $C_1(\mathbf{f}, \bfsigma)$, $C_2(\mathbf{f}, \bfsigma)$ are defined in (\ref{C_1}) and (\ref{C_2}).
\end{lem}
\textbf{Proof:} For simplicity, we consider $L+2\leq i_{\star} \leq n-L-1$, this doesn't sacrifice any generality because we can always rotate the index to make it true. So $B_{i_{\star}} = \{ i_{\star}-L, ..., i_{\star}+L\}$.

Suppose $j \in B^c_{i_{\star}}$, since $\mb{x}^{\text{l}}_j(t) =  \mb{x}^{\text{o}}_j(t)$ and $d(j,i_{\star})\geq L+1$, (\ref{xlp2}) follows from the conclusion of Lemma \ref{bound_opz_l}.

Suppose $j \in B_{i_{\star}}$, we observe that both  $\mb{x}_j^{\text{l}}(t)$ and $\mb{x}_j^{\text{p}}(t)$ follow the evolutionary system
\begin{align*}
&d\mb{x}_j(t) =\mathbf{f}_{j}(t,\mb{x}_{j-1}(t), \mb{x}_j(t),\mb{x}_{j+1}(t) )dt  \\
&\qquad \qquad+ \bfsigma_{j}(t,\mb{x}_j(t))d\bfW_j(t), \ t \in [0,T],
\end{align*}
with initial value $(\mb x^{\text{o}}_{i_{\star}-L}, ...,\mb x^{\text{o}}_{i_{\star}-1}, \mb x^{\text{p}}_{i_{\star}}, \mb x^{\text{o}}_{i_{\star}+1},..., \mb x^{\text{o}}_{i_{\star}+L} )^{T}$. But when $j\in B_{i_{\star}}^{c}$, it is restricted that $\mb{x}_{j}^{\text{l}}(t) = \mb{x}_{j}^{\text{o}}(t)$, where $\mb{x}^{\text{o}}(t)$ is the solution of (\ref{sys:original}) with initial value $(\mb{x}_1^{\text{o}}, ..., \mb{x}_m^{\text{o}})^{T}$. Recall that $\mb{x}^{\text{p}}(t)$ also solves (\ref{sys:original}), but with a locally perturbed initial value, which can be written as $(\mb{x}_1^{\text{o}}, ..., \mb{x}_{i_{\star-1}}^{\text{o}}, \mb{x}_{i_\star}^{\text{p}}, \mb{x}_{i_{\star}+1}^{\text{o}},..., \mb{x}_m^{\text{o}})^{T}$. Thus, for $i_{\star}-L \leq j \leq i_{\star}+L$, we have
\begin{align*}
d(\bfx^{\text{l}}_j(t) - \bfx^{\text{p}}_j(t)) = \Delta_{\mathbf{f}_j}(t) dt + \Delta_{\bm\sigma_{j}}(t)d\bfW_j(t),
\end{align*} 
with 
\begin{align*}
& \Delta_{\mathbf{f}_j}(t)  = \mathbf{f}_j(t, \bfx^{\text{l}}_{j-1}(t), \bfx^{\text{l}}_j(t),\bfx^{\text{l}}_{j+1}(t) ) \\
& \qquad \qquad- \mathbf{f}_j(t, \bfx^{\text{p}}_{j-1}(t), \bfx^{\text{p}}_j(t),\bfx^{\text{p}}_{j+1}(t) ),\\
&  \Delta_{\bm\sigma_{j}}(t) = (\bm{\sigma}_j(t,\bfx^{\text{l}}_j(t))- \bm{\sigma}_j(t,\bfx^{\text{p}}_j(t))).
\end{align*}
According to It$\hat{\text{o}}$'s formula,
\begin{align}\label{dxlp2}
\begin{split}
&d\|\bfx^{\text{l}}_j(t) - \bfx^{\text{p}}_j(t)\|^2\\
&= (2(\bfx^{\text{l}}_j(t) - \bfx^{\text{p}}_j(t))^{T}\Delta_{\mathbf{f}_j}(t) +   \|\Delta_{\bm\sigma_{j}}(t)\|^2)dt \\
& \quad +2(\bfx^{\text{l}}_j(t) - \bfx^{\text{p}}_j(t))^{T}\Delta_{\bm\sigma_{j}}(t) d\bfW_j(t).
\end{split}
\end{align} 
The result (\ref{dxlp2}) is similar to (\ref{dxop2}), solving it \black{and} according to (\ref{Exop2}), we obtain
\begin{align}\label{Exlp2}
\begin{split}
&\E[(\bfx^{\text{l}}_j(t) - \bfx^{\text{p}}_j(t))^2] - (\bfx^{\text{l}}_j(0) - \bfx^{\text{p}}_j(0))^2 \\
& \leq  C_{\mathbf{f}}\int_{0}^{t}\E[(\bfx^{\text{l}}_{j-1}(s) - \bfx^{\text{p}}_{j-1}(s))^2]ds \\
& \quad+ (C_{\mathbf{f}} +C_{\bm\sigma} + 1) \int_{0}^{t}\E[(\bfx^{\text{l}}_j(s) - \bfx^{\text{p}}_j(s))^2]ds \\
& \quad+ C_{\mathbf{f}}  \int_{0}^{t}\E[(\bfx^{\text{l}}_{j+1}(s) - \bfx^{\text{p}}_{j+1}(s))^2]ds. 
\end{split}
\end{align}
We define an $(2L+1)\times 1$ vector $\vec{\Delta}(t) = \big(\E[\|\mb{x}^{\text{l}}_{i_{\star}-L}(t) - \mb{x}_{i_{\star}-L}^{\text{p}}(t)\|^2], ..., \E[\|\mb{x}^{\text{l}}_{i_{\star}+L}(t) - \mb{x}_{i_{\star}+L}^{\text{p}}(t)\|^2] \big)^{T}$, whose $j$-th element $\vec{\Delta}_j(t)= \E[\|\mb{x}^{\text{l}}_{i_{\star}-L+j-1}(t) - \mb{x}_{i_{\star}-L+j-1}^{\text{p}}(t)\|^2]$, and observe that $\vec{\Delta}(0) = \mathbf{0}_{(2L+1)\times 1}$. Since $\mb{x}^{\text{l}}_{i_{\star}-L-1}(t) = \mb{x}^{\text{o}}_{i_{\star}-L-1}(t)$ and $\mb{x}^{\text{l}}_{i_{\star}+L+1}(t) = \mb{x}^{\text{o}}_{i_{\star}+L+1}(t)$, we can write (\ref{Exlp2}) as the following vector form
\begin{align}\label{Delta}
\vec{\Delta}(t) \preceq \int_{0}^{t}M \cdot \vec{\Delta}(s) ds + \vec{\delta}(t),
\end{align}
where $M$ is an $(2L+1) \times (2L+1)$ tridiagonal matrix with 
\begin{align*}
& M_{1,1} = C_{\mathbf{f}}+C_{\bfsigma}+1, \ M_{1,2} =C_{\mathbf{f}}, \\
& M_{j,j} =C_{\mathbf{f}}+C_{\bfsigma}+1, \ M_{j,j+1} =M_{j,j-1} =C_{\mathbf{f}},\\
&\qquad  \ \text{for} \  2 \leq j \leq 2L  \\
&M_{2L+1,2L} =  C_{\mathbf{f}},  \ M_{2L+1,2L+1} = C_{\mathbf{f}}+C_{\bfsigma}+ 1,
\end{align*} 
and $\vec{\delta}(t)$ is an $(2L+1)$-dimensional column vector with $\vec{\delta}_{1}(t) =C_{\mathbf{f}}\int_{0}^{t} \E[\|\mb{x}^{\text{o}}_{i_{\star}-L-1}(s) - \mb{x}_{i_{\star}-L-1}^{\text{p}}(s)\|^2]ds$, $\vec{\delta}_{2L+1}(t) \\= C_{\mathbf{f}}\int_{0}^{t}\E[\|\mb{x}^{\text{o}}_{i_{\star}+L+1}(s) - \mb{x}_{i_{\star}+L+1}^{\text{p}}(s)\|^2]ds$, while other entries  are 0. According to Lemma \ref{bound_opz_l}, we have
\begin{align}\label{maxdel}
\begin{split}
&\max_{j=1,...,2L+1}{\{|\vec{\delta}_{j}(t)|\}} \\
& \leq C_{\mathbf{f}} \int_{0}^{t}\| \mb{x}^{\text{o}}_{i_\star}-\mb{x}^{\text{p}}_{i_\star} \|^2 e^{C_1(\mathbf{f}, \bfsigma)\cdot s} e^{-C_d(L+1)}ds \\
& \leq \dfrac{C_{\mathbf{f}}}{C_1(\mathbf{f}, \bfsigma)}\| \mb{x}^{\text{o}}_{i_\star}-\mb{x}^{\text{p}}_{i_\star} \|^2 e^{C_1(\mathbf{f}, \bfsigma)\cdot t} e^{-C_d(L+1)}.
\end{split}
\end{align}
By Lemma \ref{gron_s}, under (\ref{Delta}), we have 
\begin{align}\label{Deltat}
\vec{\Delta}(t) \preceq \vec{\delta}(t) + \int_{0}^{t} e^{M(t-s)}\cdot M \cdot \vec{\delta}(s)ds.
\end{align}
For the $(2L+1)\times 1$ vector $(M \cdot \vec{\delta}(t))$, we observe 
\begin{align*}
&(M \cdot \vec{\delta}(t))_{1} = (C_{\mathbf{f}} + C_{\bfsigma}+1)\vec{\delta}_{1}(t) , (M \cdot \vec{\delta}(t))_{2} =C_{\mathbf{f}}\vec{\delta}_{1}(t),\\
&(M \cdot \vec{\delta}(t))_{2L} = C_{\mathbf{f}}\vec{\delta}_{2L+1}(t) , \\
&(M \cdot \vec{\delta}(t))_{2L+1} =(C_{\mathbf{f}} + C_{\bfsigma}+1)\vec{\delta}_{2L+1}(t),
\end{align*}
while other entries are 0. Together with (\ref{maxdel}), we have 
\begin{align}\label{maxMdel}
\begin{split}
&\max_{j=1,...,2L+1}{\{|(M \cdot \vec{\delta}(t))_{j,1}|\}} \\
&\leq \dfrac{C_{\mathbf{f}}}{C_1(\mathbf{f}, \bfsigma)}(C_{\mathbf{f}} + C_{\bfsigma}+1)\| \mb{x}^{\text{o}}_{i_\star}-\mb{x}^{\text{p}}_{i_\star} \|^2 e^{C_1(\mathbf{f}, \bfsigma)\cdot t} e^{-C_d(L+1)}.
\end{split}
\end{align}
Since $\max\limits_{j,i=1,...,2L+1} M_{j,i} = (C_{\mathbf{f}} + C_{\bfsigma}+1)$, according to Lemma \ref{ineq} and the conclusion of Lemma \ref{diag} with $C=C_d$, we have
\begin{align}\label{eMt-s}
\begin{split}
|(e^{M(t-s)})_{j,i}| \leq |(e^{Mt})_{j,i}| \leq e^{C_1(\mathbf{f}, \bfsigma)t}.
\end{split}
\end{align}
From (\ref{Deltat}), together with (\ref{maxdel}), (\ref{maxMdel}) and (\ref{eMt-s}), we obtain
\begin{align*}
&\vec{\Delta}_j(t) \\
&\leq |\vec{\delta}_j(t)| +  \sum_{k=1,2,2L,2L+1 }\int_{0}^{t}|(e^{M(t-s)})_{j,k}| |(M\cdot \vec{\delta}(s))_{k,1}|ds \\
& \leq \max_{j=1,...,2L+1}{\{|\vec{\delta}_{j}(t)|\}}  \\
& \quad+  4e^{C_1(\mathbf{f}, \bfsigma)t}\int_{0}^{t}\max_{j=1,...,2L+1}{\{|(M \cdot \vec{\delta}(s))_{j,1}|\}}ds \\
& \leq \dfrac{C_{\mathbf{f}}}{C_1(\mathbf{f}, \bfsigma)}\| \mb{x}^{\text{o}}_{i_\star}-\mb{x}^{\text{p}}_{i_\star} \|^2 e^{C_1(\mathbf{f}, \bfsigma)\cdot t} e^{-C_d(L+1)} \\
&\quad \cdot  (1+ 4\int_{0}^{t}(C_{\mathbf{f}} + C_{\bfsigma}+1) e^{C_1(\mathbf{f}, \bfsigma)\cdot s}ds ) \\
& \leq \dfrac{C_{\mathbf{f}}}{C_1(\mathbf{f}, \bfsigma)}\| \mb{x}^{\text{o}}_{i_\star}-\mb{x}^{\text{p}}_{i_\star} \|^2 e^{C_1(\mathbf{f}, \bfsigma)\cdot t} e^{-C_d(L+1)} \\
& \quad \cdot (1+\dfrac{4(e^{C_1(\mathbf{f}, \bfsigma)t}-1)}{e^{C_d}+e^{-C_d}+1}) \\
& \leq \dfrac{2C_{\mathbf{f}}}{C_1(\mathbf{f}, \bfsigma)} \| \mb{x}^{\text{o}}_{i_\star}-\mb{x}^{\text{p}}_{i_\star} \|^2 e^{2C_1(\mathbf{f}, \bfsigma)\cdot t} e^{-C_d(L+1)},
\end{align*}
where the last inequality is derived by using $e^{C_d}+e^{-C_d}+1 \geq 2$. Recall the definition of $\vec{\Delta}_j(t)$, the proof is complete.
\hfill $\Box$

\begin{lem}\label{l_con}
	Under the same settings in Lemma \ref{bound_lpz}, for any given $\epsilon>0$, if the local domain radius $L$ satisfies
	\begin{align*}
	L \geq \dfrac{\log{\Big( \dfrac{\epsilon}{C_2(\mathbf{f}, \bfsigma) \|\mb{x}^{\emph{o}}_{i_\star}-\mb{x}^{\emph{p}}_{i_\star}\|^2} \Big)}}{-C_d}   + \dfrac{2C_1(\mathbf{f}, \bfsigma)}{C_d}\cdot T,
	\end{align*}
	then $\E[ \|\bfx^{\text{\emph{l}}}_j(t) - \mb{x}_j^{\text{\emph{p}}}(t)\|^2 ]\leq \epsilon$ for all $t\leq T$ and $j=1,...,m$. 
\end{lem}
\textbf{Proof:} According to Lemma \ref{bound_lpz}, it's equivalent for us to solve 
\begin{align*}
C_2(\mathbf{f}, \bfsigma) \cdot \| \mb{x}^{\text{o}}_{i_\star}-\mb{x}^{\text{p}}_{i_\star} \|^2 e^{2C_1(\mathbf{f}, \bfsigma)\cdot t} e^{-C_d(L+1)} \leq \epsilon
\end{align*}
for $t \in [0,T]$, which can be obtained by solving
\begin{align*}
&e^{-C_d(L+1)} \leq \dfrac{\epsilon}{C_2(\mathbf{f}, \bfsigma)\cdot \|\mb{x}^{\text{o}}_{i_\star}-\mb{x}^{\text{p}}_{i_\star}\|^2 \cdot e^{2C_1(\mathbf{f}, \bfsigma)\cdot T } }.
\end{align*}
After taking log for both sides, we obtain that above result is true if 
\begin{align*}
L \geq \dfrac{\log{\Big( \dfrac{\epsilon}{C_2(\mathbf{f}, \bfsigma) \|\mb{x}^{\text{o}}_{i_\star}-\mb{x}^{\text{p}}_{i_\star}\|^2} \Big)}}{-C_d}   + \dfrac{2C_1(\mathbf{f}, \bfsigma)}{C_d}\cdot T.
\end{align*}
The proof is finished.
\hfill $\Box$

\begin{lem}\label{bound_opz_E}
	Under Assumption \ref{lip}, with $\tilde{\mb{x}}^{\emph{o}}_j(ih)$, $\tilde{\mb{x}}^{\emph{p}}_j(ih)$, for $j=1,...,m$ and $i=0,...,T/h$, defined in Section \ref{Euler}, we have
	\begin{align}\label{lemtilxop2}
	\begin{split}
	& \| \tilde{\mb{x}}^{\emph{o}}_j(ih) - \tilde{\mb{x}}^{\emph{p}}_j(ih) \|^2\\
	&\leq  e^{-C_d\cdot d(j,i_{\star})}\cdot e^{C_1(\mathbf{f}, \bfsigma)( 1+h)ih} \|\mb{x}^{\emph{o}}_{i_\star}-\mb{x}^{\emph{p}}_{i_\star}\|^2,
	\end{split}
	\end{align}
	where $C_d$ is any given positive constant, $C_1(\mathbf{f}, \bfsigma)$ is defined in (\ref{C_1}).
\end{lem}
\textbf{Proof:} Since $\bfxtildeo_{j}(ih)$ are obtained by iterating 
\begin{align}\label{tilxo}
\begin{split}
\bfxtildeo_j((i+1)h)&=  \bfxtildeo_j(ih)  +\bfsigma_j(ih,\bfxtildeo_j(ih))\sqrt{h}W_{i,j} \\
& \quad  + \mathbf{f}_j(ih,\bfxtildeo_{j-1}(ih), \bfxtildeo_j(ih),\bfxtildeo_{j+1}(ih) )h,
\end{split}
\end{align}
with initial value $\bfxtildeo_{j}(0)=\bfxo_{j}$. Comparing it with (\ref{tilxp}), we have 
\begin{align}\label{tilxop}
\begin{split}
&\bfxtildeo_j((i+1)h) - \bfxtildep_j((i+1)h)\\
&=  \bfxtildeo_j(ih) - \bfxtildep_j(ih) +\tilde{\Delta}_{\bfsigma_j} (ih) \sqrt{h}W_{i,j} + \tilde{\Delta}_{\mathbf{f}_j} (ih) h,
\end{split}
\end{align}
with
\begin{align*}
\tilde{\Delta}_{\bfsigma_j} (ih) &= \bfsigma_j(ih,\bfxtildeo_j(ih)) - \bfsigma_j(ih,\bfxtildep_j(ih)), \\
\tilde{\Delta}_{\mathbf{f}_j} (ih) & = \mathbf{f}_j(ih,\bfxtildeo_{j-1}(ih), \bfxtildeo_j(ih),\bfxtildeo_{j+1}(ih) ) \\
&\quad - \mathbf{f}_j(ih,\bfxtildep_{j-1}(ih), \bfxtildep_j(ih),\bfxtildep_{j+1}(ih) ).
\end{align*}
Since $\tilde{\mb{x}}_j^{\text{o}}(ih)$ and $\tilde{\mb{x}}_j^{\text{p}}(ih)$ are independent with $W_{i,j}$, together with Assumption \ref{lip}, and Cauchy-Schwartz inequality, from (\ref{tilxop}), we have
\begin{align}\label{Etilxop2}
\begin{split}
& \E[\|\tilde{\mb{x}}_j^{\text{o}}((i+1)h) - \tilde{\mb{x}}_j^{\text{p}}((i+1)h)\|^2] \\
&= \E[\|\tilde{\mb{x}}_j^{\text{o}}(ih) - \tilde{\mb{x}}_j^{\text{p}}(ih) \|^2] +  \E[\|\tilde{\Delta}_{\bfsigma_j} (ih) \sqrt{h}W_{i,j} \|^2] \\
& \quad + \E[\|\tilde{\Delta}_{\mathbf{f}_j} (ih) h\|^2]  + 2\E[(\tilde{\mb{x}}_j^{\text{o}}(ih) - \tilde{\mb{x}}_j^{\text{p}}(ih))^{T} \tilde{\Delta}_{\mathbf{f}_j} (ih) h] \\
& \leq \E[\|\tilde{\mb{x}}_j^{\text{o}}(ih) - \tilde{\mb{x}}_j^{\text{p}}(ih) \|^2] +   \E[\|\tilde{\Delta}_{\bfsigma_j} (ih)\|^2] h\\
& \quad + \E[\|\tilde{\Delta}_{\mathbf{f}_j} (ih) \|^2] h^2 \\
&\quad + ( \E[\|\tilde{\mb{x}}_j^{\text{o}}(ih) - \tilde{\mb{x}}_j^{\text{p}}(ih) \|^2] + \E[\|\tilde{\Delta}_{\mathbf{f}_j} (ih) \|^2] )h \\
&\leq (C_{\mathbf{f}}h+C_{\mathbf{f}}h^2 )\E[(\tilde{\mb{x}}^{\text{o}}_{j-1}(ih) - \tilde{\mb{x}}^{\text{p}}_{j-1}(ih) )^2]  +\\
&\quad (1+(C_{\bfsigma} + C_{\mathbf{f}}+1)h +C_{\mathbf{f}}h^2) \E[(\tilde{\mb{x}}_j^{\text{o}}(ih) - \tilde{\mb{x}}_j^{\text{p}}(ih) )^2] \\
& \quad +(C_{\mathbf{f}}h+C_{\mathbf{f}}h^2 ) \E[\|\tilde{\mb{x}}^{\text{o}}_{j+1}(ih) - \tilde{\mb{x}}^{\text{p}}_{j+1}(ih) \|^2]. 
\end{split}
\end{align}
Equivalently, we have 
\begin{align}\label{Deltai+1h}
\vec{\Delta}((i+1)h) \preceq (\mathbf{I}_{m} + M) \cdot \vec{\Delta}(ih),
\end{align}
where $\vec{\Delta}(ih)$ is defined to be an $m\times 1$ vector with its $j$-th entry $\vec{\Delta}_j(ih) = \E[\|\tilde{\mb{x}}_j^{\text{o}}(ih) - \tilde{\mb{x}}_j^{\text{p}}(ih)\|^2] $ for $i=0,...,T/h$; $M$ is an $m \times m$ matrix with 
\begin{align*}
&M_{1,1} =(C_{\bfsigma} + C_{\mathbf{f}}+1)h +C_{\mathbf{f}}h^2,  M_{1,2} =C_{\mathbf{f}}h+C_{\mathbf{f}}h^2, \\
&M_{1,m} =C_{\mathbf{f}}h+C_{\mathbf{f}}h^2, \\
&M_{j,j-1} =C_{\mathbf{f}}h+C_{\mathbf{f}}h^2, M_{j,j} =(C_{\bfsigma}+ C_{\mathbf{f}}+1)h +C_{\mathbf{f}}h^2, \\
&M_{j,j+1} = C_{\mathbf{f}}h+C_{\mathbf{f}}h^2, \quad \text{for} \ j=2,...,m-1,\\
&M_{m,1} = C_{\mathbf{f}}h+C_{\mathbf{f}}h^2,M_{m,m-1} = C_{\mathbf{f}}h+C_{\mathbf{f}}h^2, \\
& M_{m,m} =(C_{\bfsigma}+ C_{\mathbf{f}}+1)h +C_{\mathbf{f}}h^2,
\end{align*}
and other entries are 0. After iterating (\ref{Deltai+1h}) for $i$ times, we have
\begin{align}\label{Deltaih}
\vec{\Delta}(ih) \preceq (\mathbf{I}_{m} + M)^{i} \cdot \vec{\Delta}(0), 
\end{align}
and for $\vec{\Delta}(0)$, we know $\vec{\Delta}_{i_{\star}}(0) = \|\mb{x}^{\text{o}}_{i_\star}-\mb{x}^{\text{p}}_{i_\star}\|^2$ and other entries are 0. We observe 
\begin{align}\label{IMe^M}
\mathbf{I}_{m}+M \preceq e^{M},
\end{align}
together with Lemma \ref{diag}, $\max\limits_{j,k = 1,2,...,m} |(iM)_{j,k}| = ((C_{\bfsigma}+C_{\mathbf{f}}+1) +  C_{\mathbf{f}}h)ih$ and (\ref{M^k}), we have, for $j=1,...,m$,
\begin{align}\label{1+M^i}
\begin{split}
|((\mathbf{I}_{m}+ M)^{i})_{j,i_{\star}}| & \leq |((e^{M})^i)_{j,i_{\star}}| = |(e^{iM})_{j,i_{\star}}| \\
&\leq e^{-C_d\cdot d(j,i_{\star})}\cdot e^{C_1(\mathbf{f}, \bfsigma) (1+h)ih}.
\end{split}
\end{align}
Substituting (\ref{1+M^i}) into (\ref{Deltaih}) results in (\ref{lemtilxop2}).
\hfill $\Box$

\begin{lem}\label{bound_lpz_E}
	Under Assumption \ref{lip}, with $\tilde{\mb{x}}^{\emph{o}}_j(ih)$ $\tilde{\mb{x}}^{\emph{l}}_j(ih)$, $\tilde{\mb{x}}^{\emph{p}}_j(ih)$, for $j=1,...,m$ and $i=0,...,T/h$, defined in Section \ref{Euler}, we have
	\begin{align}
	\begin{split}
	& \E[\|\tilde{\mb{x}}^{\text{\emph{l}}}_j(ih) - \tilde{\mb{x}}_j^{\text{\emph{p}}}(ih)\|^2] \\
	&\leq C_2(\mathbf{f}, \bfsigma) e^{2C_1(\mathbf{f}, \bfsigma) (1+h)ih} e^{-C_d(L+1)}  \|\mb{x}^{\emph{o}}_{i_\star}-\mb{x}^{\emph{p}}_{i_\star}\|^2,
	\end{split}
	\end{align}
	where $C_d$ is any given positive constant, $C_1(\mathbf{f}, \bfsigma)$ and $C_2(\mathbf{f}, \bfsigma)$ are defined in (\ref{C_1}) and (\ref{C_2}).
\end{lem}
\textbf{Proof:} Suppose $j \in B^c_{i_{\star}}$, since $d(j,i_{\star})\geq L+1$, and $\tilde{\mb{x}}^{\text{l}}_j(ih) = \tilde{\mb{x}}^{\text{o}}_j(ih) $, the result follows from Lemma \ref{bound_opz_E}.

Suppose $j \in B_{i_{\star}} $, comparing (\ref{tilxp}) and (\ref{eqn:EMloc}), we have
\begin{align}\label{tilxlp}
\begin{split}
&\bfxtildel_j((i+1)h) - \bfxtildep_j((i+1)h)\\
&=  \bfxtildel_j(ih) - \bfxtildep_j(ih) +\tilde{\Delta}_{\bfsigma_j} (ih) \sqrt{h}W_{i,j} + \tilde{\Delta}_{\mathbf{f}_j} (ih) h,
\end{split}
\end{align}
with
\begin{align*}
\tilde{\Delta}_{\bfsigma_j} (ih) &= \bfsigma_j(ih,\bfxtildel_j(ih)) - \bfsigma_j(ih,\bfxtildep_j(ih)), \\
\tilde{\Delta}_{\mathbf{f}_j} (ih) & = \mathbf{f}_j(ih,\bfxtildel_{j-1}(ih), \bfxtildel_j(ih),\bfxtildel_{j+1}(ih) ) \\
&\quad - \mathbf{f}_j(ih,\bfxtildep_{j-1}(ih), \bfxtildep_j(ih),\bfxtildep_{j+1}(ih) ).
\end{align*} 
And it's required that $\tilde{\mb{x}}_{k}^{\text{l}}(ih) = \tilde{\mb{x}}_{k}^{\text{o}}(ih)$ for $k\in B_{i_{\star}}^{c}$ during the evolution of $\bfxtildel(ih)$. Since (\ref{tilxlp}) is similar to (\ref{tilxop}), according to (\ref{Etilxop2}), we have
\begin{align*}
& \E[\|\tilde{\mb{x}}_j^{\text{l}}((i+1)h) - \tilde{\mb{x}}_j^{\text{p}}((i+1)h)\|^2] \\
&\leq  (C_{\mathbf{f}}h+C_{\mathbf{f}}h^2 )\E[(\tilde{\mb{x}}^{\text{l}}_{j-1}(ih) - \tilde{\mb{x}}^{\text{p}}_{j-1}(ih) )^2]  \\
&\quad + (1+(C_{\bfsigma} + C_{\mathbf{f}}+1)h +C_{\mathbf{f}}h^2) \E[(\tilde{\mb{x}}_j^{\text{l}}(ih) - \tilde{\mb{x}}_j^{\text{p}}(ih) )^2] \\
&\quad +(C_{\mathbf{f}}h+C_{\mathbf{f}}h^2 ) \E[\|\tilde{\mb{x}}^{\text{l}}_{j+1}(ih) - \tilde{\mb{x}}^{\text{p}}_{j+1}(ih) \|^2]. 
\end{align*}
Namely, for the $(2L+1) \times 1$ vector $\vec{\Delta}(ih)$ whose $j$-th entry $\vec{\Delta}_j(ih) = \E[ \|\tilde{\mb{x}}^{\text{l}}_{i_{\star}-L+j-1}(ih) - \tilde{\mb{x}}^{\text{p}}_{i_{\star}-L+j-1}(ih)\|^2]$ for $j=1,...,2L+1$, we have 
\begin{align}\label{Deltai+1}
\vec{\Delta}((i+1)h) \preceq (\mathbf{I}_{(2L+1)} + M) \cdot \vec{\Delta}(ih) + \vec{\delta}(ih),
\end{align}
where $M$ is an $(2L+1)$ by $(2L+1)$ tridiagonal matrix with 
\begin{align*}
&M_{1,1} = (C_{\bfsigma} + C_{\mathbf{f}}+1)h +C_{\mathbf{f}}h^2,  M_{1,2} =C_{\mathbf{f}}h+C_{\mathbf{f}}h^2, \\
&M_{j,j-1} =C_{\mathbf{f}}h+C_{\mathbf{f}}h^2, M_{j,j} =(C_{\bfsigma} + C_{\mathbf{f}}+1)h +C_{\mathbf{f}}h^2, \\
&M_{j,j+1} = C_{\mathbf{f}}h+C_{\mathbf{f}}h^2, \qquad \text{for} \  2 \leq j \leq 2L  \\
&M_{2L+1,2L} =  C_{\mathbf{f}}h+C_{\mathbf{f}}h^2,  \\
&M_{2L+1,2L+1} = (C_{\bfsigma} + C_{\mathbf{f}}+1)h +C_{\mathbf{f}}h^2,
\end{align*} 
and $\vec{\delta}(ih)$ is an $(2L+1)$-dimensional vector with 
\[
\vec{{\delta}}_{1}(ih) =(C_{\mathbf{f}}+C_{\mathbf{f}}h)h \E[\|\tilde{\mb{x}}^{\text{o}}_{i_{\star}-L-1}(ih) - \tilde{\mb{x}}^{\text{p}}_{i_{\star}-L-1}(ih)\|^2],
\] 
\[
\vec{\delta}_{2L+1}(ih) = (C_{\mathbf{f}}+C_{\mathbf{f}}h)h\E[\|\tilde{\mb{x}}^{\text{o}}_{i_{\star}+L+1}(ih) - \tilde{\mb{x}}_{i_{\star}+L+1}^{\text{p}}(ih)\|^2]
\]
and other entries are 0. After iterating (\ref{Deltai+1}) for $i$ times, we obtain 
\begin{align}\label{Deltai}
\begin{split}
\vec{\Delta}(ih) & \preceq \sum_{k=0}^{i-1}(\mathbf{I}_{2L+1} + M)^{i-1-k} \cdot \vec{\delta}(kh) \\
&\quad + (\mathbf{I}_{2L+1} + M)^{i}\cdot \vec{\Delta}(0),
\end{split}
\end{align}
and we see $\vec{\Delta}(0) = \mathbf{0}_{(2L+1) \times 1}$ from the definition of $\tilde{\mb{x}}^{\text{l}}_j(ih)$, $\tilde{\mb{x}}^{\text{p}}_j(ih)$. According to (\ref{IMe^M}), (\ref{M^k}), together with Lemma \ref{diag}, $\max\limits_{j,l = 1,2,...,2L+1} |(iM)_{j,k}| = ((C_{\bfsigma}+C_{\mathbf{f}}+1) +  C_{\mathbf{f}}h)ih$, for $l=1,...,2L+1$, we have
\begin{align}\label{I2L+1}
\begin{split}
& |((\mathbf{I}_{2L+1}+ M)^{i-1-k})_{j,l}| \\
& \leq |((e^{M})^{(i-1-k)})_{j,l}| = |(e^{(i-1-k)M})_{j,l}| \\
&\leq e^{-C_d\cdot d(j,l)} e^{C_1(\mathbf{f}, \bfsigma) (1+h)(i-1-k)h} \leq e^{C_1(\mathbf{f}, \bfsigma) (1+h)(i-1-k)h}.
\end{split}
\end{align}
According to Lemma (\ref{bound_opz_E}), we have 
\begin{align}\label{maxdelkh}
\begin{split}
&\max\{ |\vec{\delta}_{1}(kh)| , |\vec{\delta}_{2L+1}(kh)| \} \\
&\leq (C_{\mathbf{f}}h+C_{\mathbf{f}}h^2) e^{-C_d(L+1)} e^{C_1(\mathbf{f}, \bfsigma) (1+h)kh} \|\mb{x}^{\text{o}}_{i_\star}-\mb{x}^{\text{p}}_{i_\star}\|^2.
\end{split}
\end{align}
By applying Taylor's theorem, we have 
\begin{align}\label{sume^C}
\begin{split}
\sum_{k=0}^{i-1} e^{C_1(\mathbf{f}, \bfsigma) (1+h)kh} &= \dfrac{ e^{C_1(\mathbf{f}, \bfsigma) (1+h)ih}-1}{e^{C_1(\mathbf{f}, \bfsigma) (1+h)h}-1}\\
& \leq \dfrac{ e^{C_1(\mathbf{f}, \bfsigma) (1+h)ih}-1}{C_1(\mathbf{f}, \bfsigma) (1+h)h}.
\end{split}
\end{align}
Substituting (\ref{I2L+1}), (\ref{maxdelkh}), (\ref{sume^C}) into (\ref{Deltai}), we have, for $j=1,...,2L+1$, 
\begin{align*}
&\vec{\Delta}_j(ih) \\
& \leq \sum_{k=0}^{i-1}((\mathbf{I}_{(2L+1)} + M)^{i-1-k})_{j,1} \cdot \vec{\delta}_1(kh) \\
&+ \sum_{k=0}^{i-1}((\mathbf{I}_{(2L+1)} + M)^{i-1-k})_{j,2L+1} \cdot \vec{\delta}_{2L+1}(kh)\\
& \leq 2\cdot e^{C_1(\mathbf{f}, \bfsigma) (1+h)ih}\cdot e^{-C_d(L+1)} \cdot \|\mb{x}^{\text{o}}_{i_\star}-\mb{x}^{\text{p}}_{i_\star}\|^2\\
& \quad \cdot (C_{\mathbf{f}}h+C_{\mathbf{f}}h^2)  \cdot \sum_{k=0}^{i-1} e^{C_1(\mathbf{f}, \bfsigma) (1+h)kh} \\
& \leq \dfrac{2C_{\mathbf{f}}}{C_1(\mathbf{f}, \bfsigma)} e^{2C_1(\mathbf{f}, \bfsigma) (1+h)ih} e^{-C_d(L+1)}  \|\mb{x}^{\text{o}}_{i_\star}-\mb{x}^{\text{p}}_{i_\star}\|^2.
\end{align*}
Recall the definition of $\vec{\Delta}_j(ih)$, the proof is complete.
\hfill $\Box$

\begin{lem}\label{l_eul}
	Under the same settings in Lemma \ref{bound_lpz_E}, given any positive constant $\epsilon$, if only
	\begin{align*}
	L \geq \dfrac{\log{\Big( \dfrac{\epsilon}{C_2(\mathbf{f}, \bfsigma) \|\mathbf{x}^{\emph{o}}_{i_{\star}} - \mathbf{x}^{\emph{p}}_{i_{\star}}\|^2}   \Big)}}{-C_d}+\dfrac{2C_1(\mathbf{f}, \bfsigma) (1+h)}{C_d}T,
	\end{align*}
	we have $\E[\|{\tilde{\mb{x}}}^{\emph{l}}_j(ih) - \tilde{\mb{x}}_j^{\emph{p}}(ih)\|^2] \leq \epsilon$ for $j=1,...,m$ and $i=0,...,T/h$.
\end{lem}
\textbf{Proof:} According to Lemma \ref{bound_lpz_E}, we only need to solve 
\begin{align*}
C_2(\mathbf{f}, \bfsigma) \cdot e^{2C_1(\mathbf{f}, \bfsigma) (1+h)ih} \cdot e^{-C_d(L+1)}  \cdot \|\mb{x}^{\text{o}}_{i_\star}-\mb{x}^{\text{p}}_{i_\star}\|^2\leq \epsilon,
\end{align*}
for $i=0,...,T/h$, which can be obtained by solving
\begin{align*}
e^{-C_d(L+1)} \leq \dfrac{\epsilon}{C_2(\mathbf{f}, \bfsigma)\cdot e^{2C_1(\mathbf{f}, \bfsigma) (1+h)T} \cdot \|\bfxo_{i_{\star}} - \bfxp_{i_{\star}}\|^2}.
\end{align*}
After taking log for both sides, we have
\begin{align*}
L \geq \dfrac{\log{\Big( \dfrac{\epsilon}{C_2(\mathbf{f}, \bfsigma)\|\bfxo_{i_{\star}} - \bfxp_{i_{\star}}\|^2}   \Big)}}{-C_d}+\dfrac{2C_1(\mathbf{f}, \bfsigma) (1+h)}{C_d}T.
\end{align*}
The proof is complete.
\hfill $\Box$

\subsection*{Proofs of propositions and theorems}

\textbf{The proof of Proposition \ref{bound_opz}:} It's a direct \black{result} of Lemma \ref{bound_opz_l}. \hfill $\Box$\\
\textbf{The proof of Theorem \ref{thm_con}: } The conclusions are direct results of Lemma \ref{bound_lpz}-\ref{l_con}.  \hfill $\Box$\\
\textbf{The proof of Theorem \ref{thm_eul}: } The conclusions are direct results of Lemma \ref{bound_lpz_E}-\ref{l_eul}.  \hfill $\Box$

%

\bibliographystyle{spmpsci}      
\bibliography{liu}


\end{document}